\PassOptionsToPackage{table}{xcolor}
\documentclass[sigconf]{acmart}
\AtBeginDocument{%
  }

\copyrightyear{2026}
\acmYear{2026}
\setcopyright{cc}
\setcctype{by-nc-nd}
\acmConference[CHI '26]{Proceedings of the 2026 CHI Conference on Human Factors in Computing Systems}{April 13--17, 2026}{Barcelona, Spain}
\acmBooktitle{Proceedings of the 2026 CHI Conference on Human Factors in Computing Systems (CHI '26), April 13--17, 2026, Barcelona, Spain}
\acmPrice{}
\acmDOI{10.1145/3772318.3790721}
\acmISBN{979-8-4007-2278-3/2026/04}

\usepackage{multirow}
\usepackage{enumitem}

\definecolor{deepblue}{HTML}{4575B4}
\definecolor{mygreen}{RGB}{195,231,194}

\newcommand{\revision}{\textcolor{black}}
\newcommand{\method}{ExChart}
\newcommand{\benchmark}{ExChart-Bench}

\begin{document}

\title{Making Multimodal LLMs Reliable Chart Data Extractors: A Benchmark and Training Framework}

\author{Yuchen He}
\orcid{0009-0003-1035-4347}
\affiliation{%
  \institution{Zhejiang University}
  \city{Hangzhou}
  \state{Zhejiang}
  \country{China}}
\authornote{All authors are affiliated with or completed this work during their internship at the State Key Lab of CAD\&CG, Zhejiang University.}
\email{heyuchen@zju.edu.cn}

\author{Peizhi Ying}
\orcid{0009-0009-2494-8131}
\affiliation{%
  \institution{Zhejiang University}
  \city{Hangzhou}
  \state{Zhejiang}
  \country{China}}
\email{yingpeizhi@zju.edu.cn}

\author{Liqi Cheng}
\orcid{0009-0000-8868-5101}
\affiliation{%
  \institution{Zhejiang University}
  \city{Hangzhou}
  \state{Zhejiang}
  \country{China}}
\email{lycheecheng@zju.edu.cn}

\author{Kuilin Peng}
\orcid{0009-0003-0945-3989}
\affiliation{%
  \institution{Guangdong University of Technology}
  \city{Guangzhou}
  \state{Guangdong}
  \country{China}}
\email{kuilinpeng3@gmail.com}

\author{Yuan Tian}
\orcid{0009-0005-6089-7694}
\affiliation{%
  \institution{Zhejiang University}
  \city{Hangzhou}
  \state{Zhejiang}
  \country{China}}
\email{yuantian@zju.edu.cn}

\author{Dazhen Deng}
\orcid{0000-0002-9057-8353}
\affiliation{%
  \institution{Zhejiang University}
  \city{Hangzhou}
  \state{Zhejiang}
  \country{China}}
\authornote{Dazhen Deng is the corresponding author.}
\email{dengdazhen@zju.edu.cn}

\author{Yingcai Wu}
\orcid{0000-0002-1119-3237}
\affiliation{%
  \institution{Zhejiang University}
  \city{Hangzhou}
  \state{Zhejiang}
  \country{China}}
\email{ycwu@zju.edu.cn}

\renewcommand{\shortauthors}{He et al.}

\begin{abstract}
  Chart data extraction, which reverse-engineers data tables from chart images, is essential for reproducibility, analysis, retrieval, and redesign.
Existing interactive tools are reliable but tedious, and mixed-initiative systems, while more efficient, lack generalizability. Recent multimodal large language models (MLLMs) offer a unified interface for chart interpretation, yet their ability to extract accurate data tables, especially without visible labels, remains unclear.
We build a benchmark featuring diverse real-world charts without data labels to evaluate this capability. Results show that, while current MLLMs reliably reconstruct table structures, they struggle with precise value recovery.
To address this, we revisit chart data extraction from a human-centered perspective and argue that extraction should follow a progressive learning process similar to how people read charts.
Our training framework substantially improves numerical accuracy, achieving state-of-the-art performance with a 7B-parameter model. A user study further shows that our model effectively supports mixed-initiative workflows for reliable chart data extraction.
\end{abstract}


\begin{teaserfigure}
  \centering
  \includegraphics[width=\textwidth]{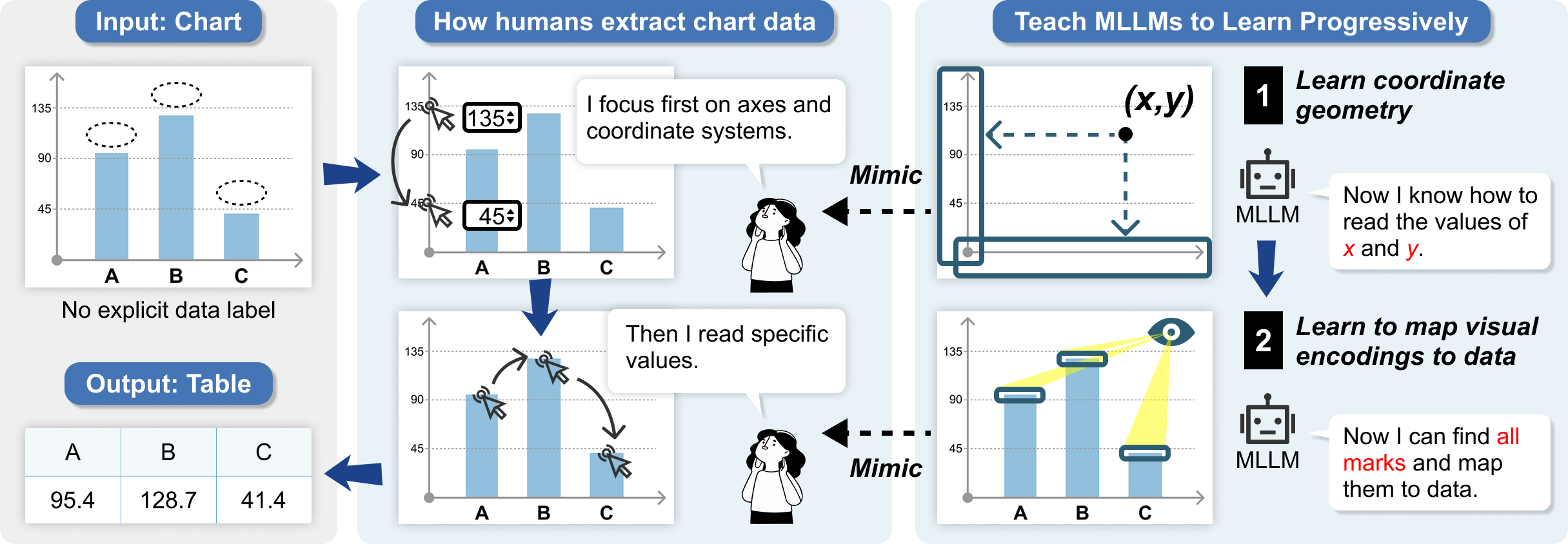}
  \caption{Chart data extraction aims to recover the underlying data table from a chart image. Although existing interactive tools and mixed-initiative workflows can produce reliable results, they fall short in terms of efficiency or generalizability. Current MLLMs are a potential solution, but they struggle to infer precise values when explicit data labels are absent. We introduce a two-stage training framework that guides MLLMs to learn in a progressive, human-like manner: first, they learn coordinate geometry; then, they map visual encodings to data.}
  \Description{This figure shows the input and output of chart data extraction. The input is a chart image, and the output is a data table. With interactive tools, users can obtain highly accurate data but need to manually calibrate and annotate the chart. This figure also shows an overview of our two-stage training framework, which mimics how humans extract data from charts.}
  \label{fig:teaser}
\end{teaserfigure}

\maketitle

\section{Introduction}

Charts serve as a primary medium for communicating quantitative information across scientific publications, business reports, and policy documents~\cite{huang2024chartsurvey,islam2019vis}. Their central role in data communication has made chart understanding and generation a vibrant research area in visualization and human-computer interaction~\cite{savva2011revision,masry2022chartqa,jung2017chartsense,deng2023visimages,tian2025chartgpt,YE2024survey,tian2025noteflow,chen2025viseval}. Among these challenges, chart data extraction, the process of recovering structured data from chart images, remains particularly valuable yet difficult. When charts are published without accompanying data tables, researchers cannot reproduce analyses, verify claims, or reuse the data for meta-analyses, redesign, or accessibility purposes~\cite{WebPlotDigitizer,chai2021crowdchart,luo2021chartocr}.

Based on efficiency and reliability, i.e., how accurately the extracted data reflects the underlying chart data~\cite{jeon2025unveiling}, existing approaches for chart data extraction fall into three categories.
Fully automatic systems~\cite{luo2021chartocr,davila2020chart,dai2018chartdecoder} such as ChartOCR~\cite{luo2021chartocr} and ChartDecoder~\cite{dai2018chartdecoder} aim to extract values without user intervention, but often struggle with real-world charts due to reliance on multi-stage parsing pipelines, hand-crafted rules, or brittle visual heuristics. These systems lack an efficient mechanism for user correction.
So far, interactive tools~\cite{WebPlotDigitizer, PlotDigitizer, DataThief}, such as WebPlotDigitizer~\cite{WebPlotDigitizer}, remain practitioners' preferred choice because they allow users to explicitly calibrate axes and select visual marks. They are reliable but tedious and time-consuming, and do not scale well to large chart collections.

To reduce manual effort while preserving user control, mixed-initiative systems~\cite{horvitz1999mixed,norman1994might,jung2017chartsense,masson2023chartdetective} combine machine learning models and interactive verification. For example, ChartSense~\cite{jung2017chartsense} integrates chart classification and mark detection into an interface that supports step-by-step refinement.
While effective in practice, these systems fundamentally rely on task-specific vision modules, i.e., detectors trained for particular mark types, axis layouts, or chart styles. Such specialized components impose strong assumptions on the input and limit generalizability to broader chart styles~\cite{jung2017chartsense,masson2023chartdetective}. Moreover, since each module handles only one subtask, the extraction workflow becomes fragmented, and users must frequently intervene at multiple stages, reducing scalability when working with large chart collections.

In contrast, recent advances in multimodal large language models (MLLMs)~\cite{wu2023multimodal,yin2024survey,vlmsurvey2024} offer a unified interface for chart interaction: a single model can answer questions, describe structures, and generate structured outputs across diverse chart types~\cite{huang2024chartsurvey,masry2022chartqa,xu2023chartbench,wang2024charxiv,hoque2022chart}.
This general-purpose capability motivates us to revisit chart data extraction from an HCI perspective: if MLLMs are becoming the interaction layer for many tasks, how should they be guided to better support reliable and scalable chart data extraction workflows?

Recent HCI work on chart interpretation~\cite{shi2025chartist} and generation~\cite{viznet,10.1145/3290605.3300358} offers an important insight here. These systems show that effective chart understanding, whether for reading or constructing visualizations, often follows a progressive cognitive process rather than a single-step mapping. People first attend to axes and coordinate systems, then recognize visual encodings, and only afterward infer specific values.
This perspective motivates us to teach MLLMs to learn chart data extraction progressively, rather than treating it as a single-step perception task. If an MLLM could internalize a similar progression, from understanding visual encodings to inferring precise numerical values, it may overcome the brittleness observed in current end-to-end approaches.
Thus, rather than directly training models to map charts to tables or code, we explore whether teaching models to understand visual encodings the way humans do, from simple to complex, can fundamentally enhance accuracy and interpretability in chart data extraction.

Guided by this human-inspired view, we develop \textbf{\method{}}, a two-stage training framework that separates foundational visual reasoning from full data extraction.
The first stage focuses on coordinate system perception. It trains the model to interpret atomic marks within specific coordinate systems, thereby reinforcing an internal geometric understanding. The second stage builds on this foundation, teaching the model to translate visual encodings into complete data tables without relying on visible labels.
This progressive approach allows MLLMs to handle unlabeled charts, a critical real-world scenario, with substantially improved stability and accuracy.

To understand where existing MLLMs fall short and to measure progress enabled by \method{}, we construct \textbf{\benchmark{}}, a benchmark spanning real-world and synthetic charts across diverse styles and encodings. A systematic evaluation of widely used MLLMs on \benchmark{} reveals a consistent bottleneck:
Most models can reproduce table structure reliably but fail to recover accurate values when data labels are absent, confirming findings from prior HCI and chart question answering studies and our practitioner interviews.
\benchmark{} helps surface these gaps and provides a principled way to evaluate both general-purpose MLLMs and specialized models trained with our proposed framework.
Across the benchmark, \method{} significantly improves Qwen2.5-VL, achieving the highest accuracy with a model of only 7B parameters. Combined with an interactive validation and correction interface, our user study further shows that MLLM-based extraction can effectively support mixed-initiative workflows, reducing manual workload without compromising reliability.
These findings suggest a promising design direction for future HCI systems: grounding MLLMs in human-like visual reasoning and integrating them into interactive workflows may achieve both efficiency and trustworthy numerical extraction at scale. Our project page
is available at \url{https://ExChart.github.io/}.

Our main contributions are as follows:
\begin{itemize}
    \item We introduce \textbf{\benchmark{}}, a benchmark for chart data extraction that contains both real-world and synthetic charts to evaluate advanced MLLMs.
    \item We propose \textbf{\method{}}, a human-inspired training framework that enhances coordinate system perception and chart-to-table extraction, significantly enhancing MLLM performance.
    \item We conduct a user study demonstrating how MLLM-based extraction supports mixed-initiative workflows, showing promise for MLLM-based chart data extraction systems.
\end{itemize}
\section{Related Work}
In this section, we review related work on chart data extraction systems and MLLM for chart understanding.

\subsection{Chart Data Extraction Systems}

Existing chart data extraction systems can be broadly categorized into two groups: fully automatic and interactive, which include mixed-initiative systems.
Fully automatic systems such as ChartDecoder~\cite{dai2018chartdecoder}, ChartOCR~\cite{luo2021chartocr}, and other computer-vision-based pipelines~\cite{davila2020chart} attempt to extract data values directly from chart images by combining OCR, object detection, and structural parsing.
However, their dependence on multi-stage pipelines and task-specific components makes them vulnerable to failures when encountering diverse chart styles, distorted scans, aesthetic variations, or charts without visible data labels. The lack of human validation and correction further limits their applicability.

Due to these limitations, interactive tools remain the preferred choice in practical use.
Interactive systems such as WebPlotDigitizer~\cite{WebPlotDigitizer}, PlotDigitizer~\cite{PlotDigitizer}, DataThief~\cite{DataThief}, and iVoLer~\cite{iVoLER} remain widely used because they support precise, user-verified extraction.
Through manual axis calibration and selection of visual marks, users maintain control over correctness.
Yet, this level of manual effort is tedious and does not scale well to large chart collections.

To reduce user effort while preserving reliability, several mixed-initiative systems~\cite{horvitz1999mixed,norman1994might} introduce automation into interactive workflows.  
For example, ChartSense~\cite{jung2017chartsense} incorporates deep learning modules for chart classification and mark detection, allowing users to correct errors rather than annotating everything from scratch.
These systems exemplify a pragmatic middle ground: automation handles low-level perception tasks, while users validate and refine results.
However, their automation components are still limited by task-specific detectors or heuristics. This restricts their ability to generalize to diverse real-world charts.
ChartDetective~\cite{masson2023chartdetective} further supports direct manipulation of chart elements and semi-automatic extraction from charts, but requires vector graphics as input.
In this work, we explore whether MLLMs can serve as a more flexible and general perception engine for data extraction. This could enable the development of efficient and reliable tools for extracting real-world chart data.

\subsection{MLLM for Chart Understanding}

With the rise of multimodal large language models (MLLMs), chart interpretation has become more flexible and conversational. A broad range of chart understanding tasks has emerged~\cite{huang2024chartsurvey}, including chart question answering (CQA)~\cite{kafle2018dvqa,ebrahimi2018figureqa,methani2020plotqa,masry2022chartqa,xu2023chartbench,masry2024chartinstruct,zhang2024tinychart,islam2024large,masry2023unichart,han2023chartllama,zeng2025cqa}, captioning or summarization~\cite{obeid2020chart,tang2023vistext,rahman2023chartsumm,islam2024large,lian2025insight,zhou2023intelligent}, chart-to-code generation~\cite{yang2024chartmimic,zhao2025chartcoder,tian2025respark}, and fact-checking~\cite{huang2023lvlms,akhtar2023chartcheck}.
These tasks require models to understand axes, marks, and other encodings, and to integrate visual understanding with natural language instructions.

A growing subset of work investigates generating structured outputs from chart images, such as chart specifications~\cite{zhao2025chartcoder,chen2024onechart,yang2024chartmimic,xu2024chartmoe,xia2023structchart}.
ChartMimic~\cite{yang2024chartmimic} and ChartCoder~\cite{zhao2025chartcoder} study translating chart images into executable specification such as Matplotlib~\cite{bisong2019matplotlib} programs, while other approaches~\cite{chen2024onechart, xia2023structchart, xu2024chartmoe, das2025chartsofthought} reconstruct chart structures or semantic representations.
These models focus on the structural interpretation of charts, such as layout and style. However, they ignore the more fundamental ability to accurately interpret the underlying data.

The chart-to-table~\cite{masry2023unichart,xu2024chartmoe,islam2024large} setting is most closely related to chart data extraction, but prior works typically assume the presence of visible data labels, where the values can be read directly from text when producing tables.
In contrast, real-world data extraction rarely includes data labels, requiring models to infer numeric values solely from geometric alignment, an ability that current MLLMs struggle to master~\cite{islam2024large,mukherjee2025encqa,chen2025misleading,wu2024chartinsights,wu2024chartinsights}.
To the best of our knowledge, no existing chart-to-table benchmarks offer diverse, real-world charts while targeting this characteristic. Without such benchmarks, we cannot determine if MLLMs can handle real-world tasks.

Improving MLLM performance for this setting also remains insufficiently studied. Existing approaches typically train models to align chart images directly with full tables~\cite{zhang2024tinychart,masry2023unichart,meng2024chartassisstant}, but we find that such end-to-end mappings struggle to recover accurate values when visible data labels are absent. We introduce a human-inspired training framework that strengthens coordinate geometry perception before learning full table extraction. This progressive strategy yields state-of-the-art accuracy on chart data extraction and provides a more reliable foundation for mixed-initiative workflows.
\section{Background}

In this section, we first present a motivating scenario that illustrates the challenges with current tools.
We then report interviews with practitioners to understand their needs and perceptions of MLLM reliability in chart data extraction.

\subsection{Motivating Scenario}

To give context to our study, we present a motivating scenario involving a researcher named Alex, who aims to analyze the relationship between review scores and acceptance rates at an academic conference. Alex has collected several charts from annual conference reports, but does not have access to the underlying datasets. To carry out the analysis, Alex must extract the numerical values encoded in these charts.

\textbf{Chart Characteristics.}
All charts are stacked bar charts. The x-axis denotes the review scores assigned to manuscripts, the y-axis shows the number of submissions, and the stacked segments represent accepted versus rejected papers. Although the charts share a similar semantic structure, their color palettes and layouts vary across years due to changes in conference branding. Importantly, none of the charts include data labels on the stacked segments.

\textbf{Trying WebPlotDigitizer.}
Because the charts have no data labels, Alex rules out OCR-based methods and begins with WebPlotDigitizer, a widely used interactive extraction tool. Alex initially tries the color-detection mode but finds the results too noisy to be usable. He switches to manual annotation, where he must calibrate axes and click each bar segment individually. The process is tedious and slow. Moreover, the extracted table requires additional post-processing: score ranges are not included in the output, and when clicking the upper segment of a stacked bar, the tool returns the cumulative height rather than the segment-specific value.
As Alex remarks, \textit{``It will take me hours to finish extracting data from all these charts, and then I still have to write scripts to clean the tables. It is too much work.''}

\textbf{Exploring Mixed-Initiative Tools.}
Alex then looks for more automatic solutions and discovers mixed-initiative systems, which combine automation with user control. However, he cannot find any tools that support his needs. For example, he finds ChartDetective appears promising, but it requires vector graphics as input, and he only has raster images. ChartSense, another system designed by HCI researchers, does not support stacked bar charts. Alex realizes that these systems are not general enough for his needs.

\textbf{ChatGPT to the Rescue?}
Still searching for a better option, Alex turns to ChatGPT with multimodal capabilities. \textit{``It can probably give me some data, but I do not know how accurate it will be,''} he thinks. He uploads a chart and prompts GPT-4o: \textit{``Extract the data from this chart and output a table.''} The model produces a table with the correct structure, but after a quick visual comparison, Alex notices several values that are clearly incorrect.
\textit{``If even I can spot these errors at a glance, how can I trust the rest?''} Alex wonders.
With no reliable automatic solution, Alex ultimately returns to WebPlotDigitizer, accepting the effort required. \textit{``I really hope there is a better way to do this,''} he sighs.

\textbf{Summary}. This scenario illustrates common challenges in real-world chart data extraction. Interactive tools, although reliable, are tedious and time-consuming. Mixed-initiative systems offer partial automation but often lack support for diverse chart types or generalize poorly. MLLMs present an appealing alternative due to their flexible, instruction-following nature, yet their reliability remains uncertain, especially when charts lack data labels.

\subsection{Interview with Practitioners}

To better understand how practitioners perceive MLLMs for chart data extraction, particularly their reliability, we then conducted interviews with five experienced practitioners (denoted as P1---P5; ages 23---28), each of whom had performed chart data extraction more than ten times in their professional workflows. Their backgrounds span finance, STEM research, and social science, offering a diverse set of perspectives.

\subsubsection{Procedure}
We conducted semi-structured interviews that encouraged participants to reflect on their past extraction workflows and to evaluate the reliability of different tools they had used. We focus on two main themes: (1) the tools they currently use, as well as the efficiency, accuracy, and usability of these tools; (2) their impressions and experiences, if any, with using MLLMs for chart data extraction.
All interviews were recorded and transcribed. We summarized the transcripts and identified recurring themes. The analysis focuses on synthesizing practitioners' perspectives on the reliability of the tools they use and MLLMs.

\subsubsection{Feedback}

\textbf{Interactive tools are reliable but tedious.}
All participants use interactive tools such as WebPlotDigitizer or Origin as their primary choice. The dominant justification was reliability: these tools offer a transparent, verifiable workflow in which each extracted value can be visually anchored to a specific mark. P5, a biomedical researcher, explained:
\textit{``I trust it because I can see exactly where each annotated point is in the chart. It is intuitive and ensures pixel-level accuracy.''}
Mark annotations were consistently described as the most time-consuming steps, especially when processing thousands of data points. All participants expressed a strong interest in more efficient solutions.

\textbf{MLLMs are convenient but not accurate.}
Four out of five participants had experimented with MLLMs such as GPT-4 or Gemini by uploading chart screenshots and requesting tables or re-rendering code, except for P5. He explained: \textit{``I once uploaded a chart and asked for a value, but the result was inaccurate, so I never tried to extract the full table.''} They appreciated the convenience of natural language prompting, but reliability was uniformly judged as low due to two issues:
\begin{enumerate}
    \item Numerical inaccuracies. Participants who tried MLLMs consistently reported encountering incorrect values in the outputs. P3 noted a common failure pattern: \textit{``When I ask the model to extract the entire table from a chart, it often repeats a value in several consecutive cells instead of the actual value.''}
    
    \item Lack of visual grounding. Participants can easily check if there is missing data in the output, but cannot verify whether the predicted values are accurate. P2 remarked: \textit{``If the model were accurate, this would not matter. But right now, I cannot confirm whether the values are correct.''}
\end{enumerate}

\textbf{Reliability is non-negotiable.}
Participants emphasized that reliability determines whether a tool can be adopted in practice. The reasons fall into two categories:
\begin{enumerate}
    \item Research rigor. Currently, MLLMs are not proven reliable for data extraction in research contexts. P1, a graduate student who extracted data from charts in scientific papers for meta-analyses, noted: \textit{``We need to report how data is obtained. If the extraction method is not considered reliable, the results will be questioned.''}
    \item Downstream risk. In many applications, extracted values are used for further analysis or decision-making, and inaccuracies can lead to harmful decisions. P4, a financial analyst, noted: \textit{``A small error can lead to the wrong investment decision.''} P3 added: \textit{``We extract historical chart data to train disease progression models. The data should be accurate for the model to learn correctly.''}
\end{enumerate}

\textbf{Summary}. Despite the potential of MLLMs for efficient, unified chart data extraction, they are not perceived as reliable. Therefore, before MLLMs can be integrated into practical workflows, two challenges must be addressed: First, quantifying their unreliability in chart data extraction, which can guide future improvements and serve as proof for model reliability; second, developing techniques to improve MLLMs' accuracy and support reliable use.

\section{Preliminary Study}

To quantify the reliability of MLLMs in chart data extraction, a benchmark must systematically evaluate their capabilities and limitations. To guide the design of such a benchmark, we first decompose the task and conduct two small-scale pilot experiments to explore factors that influence model performance.

\textbf{Task Definition.}
Chart data extraction refers to recovering the numerical values encoded in a chart image. Given a chart image $I$, the goal is to produce a table $T$ that accurately reflects the values represented by the chart's visual marks. In most real-world scenarios, extracted tables are used for further analysis, so a useful extraction should satisfy two criteria:
(1) \textit{Correct Structure:} The table must contain the correct number of rows and columns with appropriate headers to remain compatible with downstream processing.
(2) \textit{Accurate Values:} Each numerical entry should closely match the values encoded in the chart to maintain analytical validity.
These criteria reflect two levels of understanding: structure recovery tests whether the model can parse the chart layout, while value accuracy tests whether the model can map visual encodings to precise numbers. They also differ in visibility. In practice, users can easily spot errors in the table structure, but errors in the numerical values are much harder to detect. Therefore, we design two pilot experiments to separately study each aspect.

\subsection{RQ1: Can MLLMs Generate Correct Table Structure?}

To answer RQ1, we evaluate the ability of MLLMs to identify rows, columns, and headers without introducing, omitting, or misinterpreting table elements.

\textbf{Experiment Setup.}
We randomly selected 200 charts from ChartQA~\cite{masry2022chartqa}: 100 ``two-column'' charts (basic bar, basic line, pie) and 100 ``multi-column'' charts (grouped bar, stacked bar, grouped line), covering a range of chart types and different table structures.
For this experiment, we used Gemini 2.5 Flash~\cite{comanici2025gemini}, an advanced MLLM.
We chose Gemini 2.5 Flash due to its strong performance in recent VQA benchmarks, making it a representative candidate for evaluating the current capabilities of high-performing MLLMs.

Each chart was shown to the model with the prompt: \textit{``Extract the data from this chart and output it in CSV format.''} 
The generated tables were then manually evaluated against the ground truth. 
We categorized the outputs into four groups: ``correct structure,'' ``missing rows/columns,'' ``redundant rows/columns,'' and ``misinterpreted row/column .''

\begin{table}[ht]
    \centering
    \caption{Gemini 2.5 Flash's performance on generating the correct table structure.}
    \begin{tabular}{cccc}
        \toprule
        Correct & Missing & Redundant & Misinterpreted \\
        \midrule
        \textbf{194/200 (97.0\%)} & 0 & 4/200 (2.0\%) & 2/200 (1.0\%) \\
        \bottomrule
    \end{tabular}
    \label{tab:preliminary_structure}
\end{table}

\textbf{Results.}
As \autoref{tab:preliminary_structure} shows, Gemini 2.5 Flash successfully generated the correct table structure for 97\% of charts, showing that advanced MLLMs can recognize overall chart layouts and produce well-formed tabular formats. This suggests that identifying rows, columns, and headers is largely a solved problem.
The major remaining challenge lies in accurately recovering the numerical values that populate these structures.

\subsection{RQ2: Can MLLMs Recover Accurate Values?}

To answer RQ2, we evaluate how value accuracy is affected by two factors:
(1) the presence or absence of data labels, and
(2) the prompting strategy (single value at a time vs. full table extraction).

We aim to answer the following question: (1) To what extent do data labels affect numerical accuracy? (2) Does a prompting strategy that produces longer outputs to extract all values at once cause lower accuracy?

\textbf{Data Preparation.}
\revision{We randomly sampled 50 Vega-Lite~\cite{satyanarayan2016vega} specifications from nvBench~\cite{luo2021nvbench}, with 10 instances each for basic bar, stacked bar, basic line, grouped line, and pie charts, ensuring diversity in chart types and table structure complexities.}
For each Vega-Lite specification, we edited it with Python scripts and rendered two chart versions: one with data labels and one without, resulting in 100 charts (50 with data labels and 50 without).

\textbf{Experiment Setup.}
We again used Gemini 2.5 Flash~\cite{comanici2025gemini} for this experiment.
Two prompting strategies were tested:
\begin{itemize}
    \item \textbf{Single-value prompt:} \textit{``What is the value of [X]?''}, where [X] refers to a specific data point. Each value was queried individually, reducing reasoning effort but requiring multiple prompts, which is token-inefficient.
    \item \textbf{Full-table prompt:} \textit{``Extract the data from this chart and output it in CSV format.''} We also provided the model with the corresponding CSV, in which the values are masked for the model to fill in. Prompting for a table is far more efficient for real-world use, but it needs models to handle longer output while maintaining accuracy across all values.
\end{itemize}
Applying both prompts to all 100 charts produced 200 tables.

\textbf{Evaluation Metric.}
We computed numerical accuracy using MAPE:
\begin{align}
    \text{MAPE} = \frac{1}{n} \sum_{i=1}^{n} \left| \frac{v_i - \hat{v}_i}{v_i} \right| \times 100\%, \label{eq:mape}
\end{align}
where $v_i$ is the ground truth value, $\hat{v}_i$ is the predicted value, and $n$ is the total number of values. We exclude zero values, and cap individual errors at 100\% to mitigate the influence of extreme outliers.

\begin{table}[ht]
    \centering
    \caption{Gemini 2.5 Flash's accuracy in generating data values under two pairs of conditions: charts with vs. without data labels, and single-value prompt vs. full-table prompt. The values are reported in MAPE. Lower is better.}
    \begin{tabular}{l|cc}
        \toprule
        Prompt Strategy & w/ Data Labels & w/o Data Labels \\
        \midrule
        single value at a time & 1.8\% & 7.4\% \\
        full table prompting & 1.3\% & 7.2\% \\
        \bottomrule
    \end{tabular}
    \label{tab:preliminary_value}
\end{table}

\textbf{Results.}
\autoref{tab:preliminary_value} shows clear differences across conditions.
The presence of data labels substantially improves value accuracy.
When labels are visible, the model achieves an MAPE of less than 2\%, considered good. In contrast, without data labels, the MAPE increases to over 7\%, which is high enough to undermine the reliability required for analytical tasks.
The choice of prompting strategy has only a minor effect: both the single-value prompt and the full-table prompt yield similar MAPE, with the full-table prompt even slightly better. 
This suggests that while full table prompting is more efficient in practice without accuracy trade-offs, the central bottleneck lies in the model's ability to map visual encodings to precise numbers.

\subsection{Summary}

This preliminary study offers several insights for addressing the challenges of chart data extraction with MLLMs:

\textbf{Value accuracy is the main bottleneck.}
MLLMs are good at reconstructing table structure but fail at recovering accurate values, especially when data labels are absent. This gap shows that, when we evaluate and improve MLLMs for chart data extraction, numerical accuracy should be the primary focus.

\textbf{Unlabeled charts are the critical test case.}
Removing data labels increases MAPE by roughly 5\%, which is a disaster for practical use. Therefore, when evaluating the MLLMs, we need to eliminate the effect of data labels and focus on the unlabeled chart, which reflects real-world needs.

\textbf{Full table prompting is a good practice.}
Our results show that full table prompting is efficient, token-economical, and does not reduce accuracy compared with single-value prompting. Thus, full table prompting can be a practical default for chart data extraction.
\section{\benchmark{}}

Our preliminary study showed that although MLLMs can reliably generate correct table structures, their practical adoption is constrained by value inaccuracy, especially when charts lack data labels.
Practitioners emphasized their concerns about numerical fidelity, and our pilot experiments confirmed that value accuracy remains the primary bottleneck for current MLLMs. These findings highlight the need for a systematic benchmark that can reveal such limitations and guide future development. 

A dedicated benchmark serves two key purposes. First, it provides a standardized framework to assess and compare different models, offering evidence of their reliability in chart data extraction. Second, it drives progress by surfacing performance bottlenecks and informing the design of more accurate and robust methods.

\begin{figure}[ht]
    \centering
    \includegraphics[width=\linewidth]{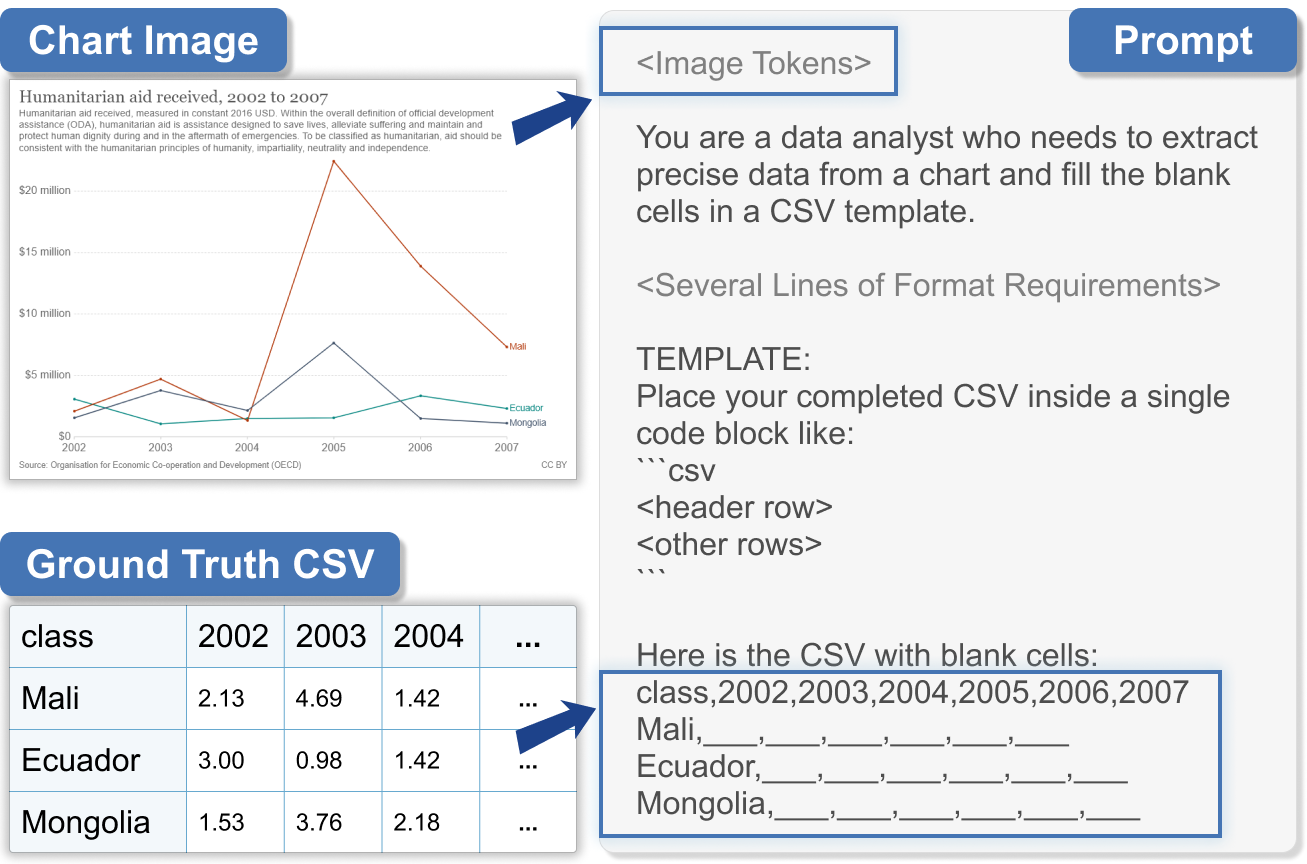}
    \caption{Prompt creation in \benchmark{}. We provide the model with a CSV template that corresponds to the chart. In the template, values are marked with placeholders.}
    \label{fig:benchmark-prompt}
    \Description{Given a chart and its corresponding CSV table, we create a prompt by replacing the values in the table with placeholders while keeping the headers and structure intact.}
\end{figure}

\subsection{Task Refinement}

To more effectively assess the core capability of MLLMs in chart data extraction, we refine the task with a focused formulation.

Given a chart image $I$ without data labels and a prompt $P_{temp}$ containing a corresponding CSV template with placeholders (as illustrated in~\autoref{fig:benchmark-prompt}), where numerical values are masked but headers and non-value cells are provided, the objective is to generate a completed table $T$ by filling in the placeholders with accurate numerical values.
This refined task isolates the model's ability to interpret visual encodings and infer precise values, an essential requirement in practical chart data extraction and the central bottleneck identified in previous studies.

\begin{figure*}[t]
    \centering
    \includegraphics[width=\linewidth]{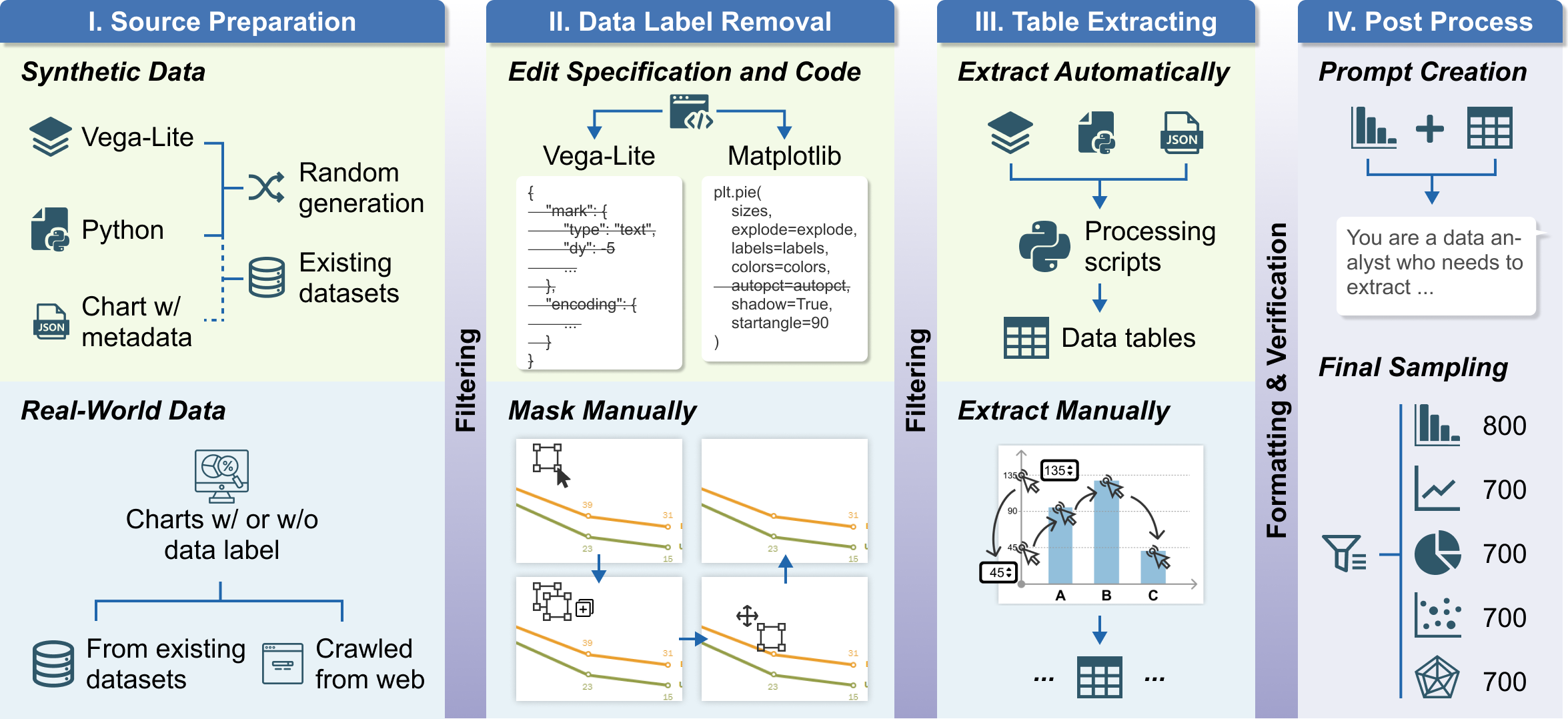}
    \caption{The benchmark dataset construction pipeline. We collected both real-world and synthetic chart sources, removed data labels, extracted data tables, and created prompts, resulting in a dataset of chart-CSV-prompt triplets.}
    \label{fig:benchmark-construction}
    \Description{The pipeline consists of several steps: source collection, source filtering, label removing, chart rendering, data table extraction, table formatting and checking, prompt creation, and final sampling of the benchmark dataset construction pipeline. We collected both real-world and synthetic chart sources, removed data labels, extracted data tables, created prompts with CSV templates, resulting in a dataset of chart-image, CSV-table, and prompt triplets.}
\end{figure*}

\subsection{Dataset Construction Pipeline}

Based on the refined task definition, we then construct the \benchmark{}.
Each sample in the final dataset includes a chart image without data labels, a ground truth CSV data table, and a prompt containing the corresponding CSV template.
\autoref{fig:benchmark-construction} illustrates the overall construction pipeline, which consists of the following steps:

\textbf{Source Preparation.}
We collect both real-world and synthetic charts to ensure diversity across chart types and visual complexities.
For real-world charts, we source images from publicly available datasets such as ChartQAPro~\cite{masry2025chartqapro} and CharXiv~\cite{wang2024charxiv}, as well as additional charts crawled from the web.
For synthetic charts, we gathered Vega-Lite specifications from existing collections, the online Vega-Lite gallery~\cite{vegalitegallery}, and Python scripts from prior datasets.
We also include charts from datasets that provide synthetic images with associated JSON metadata~\cite{zhao2025chartcoder,chen2024onechart}.  
After collecting a large set of sources, we filter out unusable samples. We remove scripts and specifications that cannot be correctly rendered.
For image-metadata pairs, we use Qwen2.5-VL 32B to identify samples with missing metadata or visible data labels and filter them out.

\textbf{Diversity Expansion.}
To further broaden coverage, we generate additional charts derived from existing Matplotlib and Vega-Lite specifications.
For style diversity, we randomly vary supported attributes such as color palettes, fonts, gridline visibility, and tick formatting.
For data diversity, we apply transformations including scaling, shifting, and adding noise to the underlying datasets.
We also modify chart configurations, for example, converting grouped bar charts into stacked versions, to enrich structural variety.

\textbf{Data Label Removal.}
For real-world charts, we manually remove all visible data labels using image editing tools.
We first mask the labels with background color and then carefully recover occluded regions by cloning surrounding pixels.
For synthetic sources, such as Python scripts or Vega/Vega-Lite specifications, we write Python scripts to remove label-related configurations.

\textbf{Synthetic Chart Augmentation.}
To simulate imperfections commonly observed in scanned or low-quality images, we apply several augmentation techniques to rendered synthetic charts.
These include small random rotations (within $\pm 5^\circ$), Gaussian noise, sharpening, contrast adjustments, ink-saving effects, and brightness variation.
We also apply downscaling to mimic low-resolution charts and use grayscale conversion or color inversion to simulate monochrome or unconventional color schemes.
These augmentations are randomly applied to all synthetic charts.

\textbf{Data Table Extraction.}
To obtain the ground truth CSV tables, we used different approaches for the synthetic and real-world charts.  
For real-world charts, we manually annotate data points using WebPlotDigitizer~\cite{WebPlotDigitizer} for maximum precision.
For synthetic charts, we write Python scripts to extract data directly from original scripts, specifications, or metadata.
We use Qwen2.5-VL 32B to check for consistency between the extracted tables and the charts and correct any discrepancies manually.

\textbf{Table Formatting and Verification.}
We standardize table formats within each chart type to ensure structural consistency. For each chart type, we select a canonical format for each chart type that results in a minimal number of cells to save tokens, and we convert all tables accordingly while preserving all information. Unit usage, delimiter styles, and column ordering are harmonized automatically using Python scripts.

\textbf{Prompt Creation.}
We design a prompt template as shown in \autoref{fig:benchmark-prompt}, which clearly defines the task and provides the model with structured input.
Each prompt includes an instruction followed by a CSV template that corresponds to the chart. In this template, all numeric values are replaced with placeholders, and the headers and structure are retained.
We also instruct the model to output its answer in a code block for easier parsing.

\textbf{Final Sampling.}
To avoid unresolvable ambiguities, we remove charts with excessive mark overlap, which makes estimating values infeasible.
Then, we sample a balanced subset of charts across types, ensuring diversity in both type and style in the final benchmark.
Since bar charts comprise a significantly larger proportion of the collected data and have more variants (e.g., stacked, grouped, vertical, and horizontal), we sample 800 bar charts and 700 charts from each of the other categories.

\begin{table}[ht]
\caption{\benchmark{} overview.}
\label{tab:benchmark-overview}
\centering
\resizebox{\linewidth}{!}{
\begin{tabular}{c|ccccc|cc}
\toprule
\multirow{2}{*}{\textbf{Total}} & \multicolumn{5}{c|}{\textbf{By Chart Type}} & \multicolumn{2}{c}{\textbf{By Source}} \\ \cmidrule(lr){2-8}
 & Bar & Line & Scatter & Pie & Radar & Real & Synthetic \\ \midrule
\multicolumn{8}{c}{\textbf{Chart Number}} \\ \midrule
3,600 & 800 & 700 & 700 & 700 & 700 & 744 & 2,856 \\ \midrule
\multicolumn{8}{c}{\textbf{Value Number}} \\ \midrule
33,757 & 9,500 & 9,494 & 8,414 & 3,055 & 3,294 & 9,907 & 23,850 \\ \bottomrule
\end{tabular}}
\end{table}

\begin{figure*}[t]
    \centering
    \includegraphics[width=\linewidth]{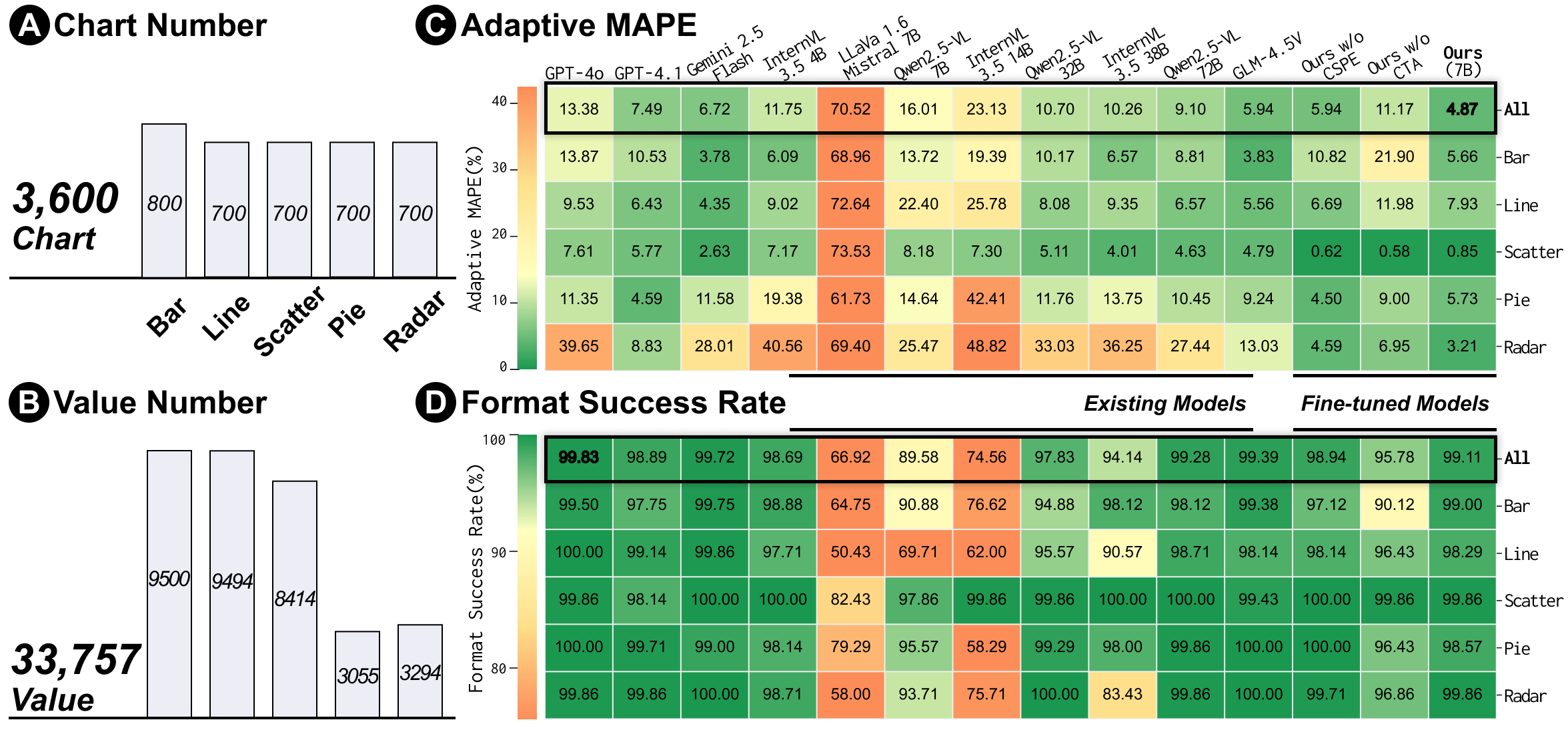}
    \caption{Distribution of benchmark samples and evaluation results of MLLMs across different chart types. (A) Distribution of chart types in the benchmark. (B) Distribution of numerical values in the benchmark. (C\&D) Adaptive MAPE and format success rate of various MLLMs on the benchmark.}
    \label{fig:benchmark-type}
    \Description{This figure shows the distribution of benchmark samples, including bar, line, scatter, pie, and radar charts. It also presents the evaluation results of various MLLMs across these chart types, including Adaptive MAPE and format success rate.}
\end{figure*}

\subsection{Benchmark Overview}
\benchmark{} contains 3,600 chart-table-prompt pairs spanning five common chart types: bar (basic, grouped, stacked), line (basic, grouped), scatter, pie, and radar charts.
Among these, 744 are real-world charts, and 2,856 are synthetic charts. In total, the dataset includes 33,757 numerical values. \autoref{tab:benchmark-overview} summarizes the dataset statistics.
To support the analysis of model performance with respect to table size, we categorize the charts into two groups: smaller and larger tables.
For each chart type, we sort the charts by number of values and split them to balance the total number of values in the ``smaller'' and ``larger'' subsets.

\subsection{Evaluation Metrics}

To evaluate model outputs, we first parse the generated text into CSV tables.
Since the models are instructed to output tables within code blocks, we begin by extracting the content enclosed by the code block fences.
The extracted text is then parsed as a serialized CSV using Python scripts.
For each value cell, we apply regular expressions to extract the numerical values, discarding any additional text or formatting.

\textbf{Format Success Rate.}
If a model fails to produce a parsable CSV or outputs a table whose shape does not match the provided CSV template, we consider it a format failure.
This metric reflects the model's ability to follow output formatting instructions.
Even with a template provided, some models still produce incorrect table shapes, suggesting that without such guidance, their ability to output a correct data table would be even more limited.

\textbf{Adaptive Mean Absolute Percentage Error.}
To evaluate numerical accuracy, we define Adaptive Mean Absolute Percentage Error (Adaptive MAPE), a variant of MAPE designed to better reflect perceptual error in chart data extraction tasks.
Unlike standard MAPE or sMAPE, which divide by the ground truth or average of predicted and ground truth values, Adaptive MAPE divides the absolute error by the maximum absolute value in the chart.
This adjustment reduces the disproportionate impact of small values on the overall error.
For instance, predicting 0.1 instead of 0.001 in a chart ranging from 0 to 100 results in a large MAPE, despite being visually indistinguishable.

Formally, for each chart, let $V_{\text{max}}$ denote the maximum absolute ground truth value. Given predicted values $\hat{v}_i$ and ground truth values $v_i$ over $n$ cells, Adaptive MAPE is defined as:
\begin{align}
    \text{Adaptive MAPE} = \frac{1}{n} \sum_{i=1}^{n} \left | \frac{\hat{v}_i - v_i}{V_{\text{max}}} \right | \times 100\%
\end{align}
For any individual error above 100\%, we cap it at 100\% to mitigate the effect of extreme outliers.
Unparsable values (e.g., non-numeric text or empty cells) are treated as having 100\% error.
For a fair comparison, values from unparsable output or shape mismatch tables are also treated as having a 100\% error.
By normalizing to the global scale of the chart rather than local value magnitudes, Adaptive MAPE provides a more perceptually meaningful and robust measure of accuracy.

\begin{table*}[t]
\caption{Model performance across different chart types. Format Success Rate measures the ability to follow instructions and output correct table structures; the higher the better. Adaptive MAPE measures overall accuracy; the lower the better.}
\label{tab:benchmark-type}
\centering
\begin{tabular}{l|c|ccccc|c|ccccc}
\toprule
\multirow{2}{*}{\textbf{Model}} & \multicolumn{6}{c|}{\textbf{Format Success Rate (\%) $\uparrow$}} & \multicolumn{6}{c}{\textbf{Adaptive MAPE (\%) $\downarrow$}} \\ \cmidrule(lr){2-13}
 & All & Bar & Line & Scatter & Pie & Radar & All & Bar & Line & Scatter & Pie & Radar\\ \midrule

\multicolumn{13}{l}{\textit{Commercial proprietary models}} \\

GPT-4o            & \textbf{99.83} & 99.50 & 100.00 & 99.86 & 100.00 & 99.86   & 13.38 & 13.87 & 9.53 & 7.61 & 11.35 & 39.65\\
GPT-4.1           & 98.89 & 97.75 & 99.14 & 98.14 & 99.71 & 99.86   & 7.49 & 10.53 & 6.43 & 5.77 & 4.59 & 8.83\\
Gemini 2.5 Flash  & 99.72 & 99.75 & 99.86 & 100.00 & 99.00 & 100.00   & 6.72 & 3.78 & 4.35 & 2.63 & 11.58 & 28.01\\ \midrule

\multicolumn{13}{l}{\textit{Open-source models (ordered by model size)}} \\

InternVL3.5 4B  & 98.69 & 98.88 & 97.71 & 100.00 & 98.14 & 98.71   & 11.75 & 6.09 & 9.02 & 7.17 & 19.38 & 40.56\\
LLaVA 1.6 Mistral 7B & 66.92 & 64.75 & 50.43 & 82.43 & 79.29 & 58.00   & 70.52 & 68.96 & 72.64 & 73.53 & 61.73 & 69.40\\
Qwen2.5-VL 7B   & 89.58 & 90.88 & 69.71 & 97.86 & 95.57 & 93.71   & 16.01 & 13.72 & 22.40 & 8.18 & 14.64 & 25.47\\
\hline
Ours w/o CSPE (7B) & 98.94 & 97.12 & 98.14 & 100.00 & 100.00 & 99.71   & 5.94 & 10.82 & 6.69 & 0.62 & 4.50 & 4.59\\
Ours w/o CTA (7B) & 95.78 & 90.12 & 96.43 & 99.86 & 96.43 & 96.86   & 11.17 & 21.90 & 11.98 & 0.58 & 9.00 & 6.95\\
\textbf{Ours (7B)} & 99.11 & 99.00 & 98.29 & 99.86 & 98.57 & 99.86   & \textbf{4.87} & 5.66 & 7.93 & 0.85 & 5.73 & 3.21\\
\hline
InternVL3.5 14B & 74.56 & 76.62 & 62.00 & 99.86 & 58.29 & 75.71   & 23.13 & 19.39 & 25.78 & 7.30 & 42.41 & 48.82\\
Qwen2.5-VL 32B  & 97.83 & 94.88 & 95.57 & 99.86 & 99.29 & 100.00   & 10.70 & 10.17 & 8.08 & 5.11 & 11.76 & 33.03\\
InternVL3.5 38B & 94.14 & 98.12 & 90.57 & 100.00 & 98.00 & 83.43   & 10.26 & 6.57 & 9.35 & 4.01 & 13.75 & 36.25\\
Qwen2.5-VL 72B  & 99.28 & 98.12 & 98.71 & 100.00 & 99.86 & 99.86   & 9.10 & 8.81 & 6.57 & 4.63 & 10.45 & 27.44\\
GLM-4.5V (>100B)        & 99.39 & 99.38 & 98.14 & 99.43 & 100.00 & 100.00   & 5.94 & 3.83 & 5.56 & 4.79 & 9.24 & 13.03\\
\bottomrule
\end{tabular}
\end{table*}

\begin{table}[t]
\caption{Impact of table size on model performance. Entries highlighted in green indicate the better result in each pair; metrics mirror~\autoref{tab:benchmark-type}.}
\label{tab:benchmark-size}
\centering
\resizebox{\linewidth}{!}{
\begin{tabular}{l|cc|cc}
\toprule
\multirow{2}{*}{\textbf{Model}} & \multicolumn{2}{c|}{\textbf{Format Success Rate (\%) $\uparrow$}} & \multicolumn{2}{c}{\textbf{Adaptive MAPE (\%) $\downarrow$}} \\ \cmidrule(lr){2-5}
 & \makebox[1.6cm][c]{Smaller} & Larger & \makebox[1.3cm][c]{Smaller} & Larger \\ \midrule
\multicolumn{5}{l}{Commercial proprietary models} \\
GPT-4o & 99.83 & \colorbox{mygreen}{99.84} & \colorbox{mygreen}{12.60} & 14.16 \\
GPT-4.1 & \colorbox{mygreen}{98.93} & 98.82 & \colorbox{mygreen}{6.37} & 8.60 \\
Gemini 2.5 Flash & \colorbox{mygreen}{99.83} & 99.53 & \colorbox{mygreen}{6.44} & 7.01 \\ \midrule
\multicolumn{5}{l}{Open-source models} \\
Qwen2.5-VL 7B & 86.29 & \colorbox{mygreen}{95.60} & 18.74 & \colorbox{mygreen}{13.29} \\
Qwen2.5-VL 32B & 97.38 & \colorbox{mygreen}{98.67} & \colorbox{mygreen}{10.48} & 10.91 \\
Qwen2.5-VL 72B & \colorbox{mygreen}{99.57} & 98.74 & \colorbox{mygreen}{6.90} & 11.30 \\
InternVL3.5 4B & \colorbox{mygreen}{99.05} & 98.04 & \colorbox{mygreen}{10.35} & 13.14 \\
InternVL3.5 14B & 65.91 & \colorbox{mygreen}{90.35} & 30.08 & \colorbox{mygreen}{16.20} \\
InternVL3.5 38B & \colorbox{mygreen}{94.45} & 93.56 & \colorbox{mygreen}{9.57} & 10.95 \\
LLaVA 1.6 Mistral 7B & 64.66 & \colorbox{mygreen}{71.04} & \colorbox{mygreen}{67.73} & 73.30 \\
GLM-4.5V & \colorbox{mygreen}{99.53} & 99.14 & \colorbox{mygreen}{5.04} & 6.84 \\ \bottomrule
\end{tabular}}
\end{table}

\subsection{Overall Analysis}

We evaluate a wide range of advanced MLLMs, including commercial proprietary models such as GPT-4o~\cite{hurst2024gpt}, GPT-4.1~\cite{gpt4_1}, and Gemini 2.5 Flash~\cite{comanici2025gemini}, as well as open-source models of various scales, including the LLaVA-Next~\cite{liu2023visual,liu2024llavanext,liu2023improvedllava}, Qwen~\cite{bai2025qwen2}, GLM~\cite{vteam2025glm45v}, and InternVL~\cite{chen2024internvl} families.
The performance across different chart types and table sizes is summarized in~\autoref{tab:benchmark-type} and~\autoref{tab:benchmark-size}, respectively.

Our analysis reveals several key findings:

\textbf{F1: None of these MLLMs are ready for chart data extraction.}
Even the best-performing model, GLM-4.5V, achieves a 99.39\% format success rate but only a 5.94\% Adaptive MAPE. Since we are using Adaptive MAPE rather than MAPE, the real percentage error for each value would be higher than 5.94\%.
This level of error can be unacceptable in domains where precision is critical.

\textbf{F2: Scaling up MLLMs on general VQA or CQA is not sufficient.}
In terms of Adaptive MAPE, GLM-4.5V, which has more than 100B parameters in total, leads all models.
Commercial proprietary models like Gemini 2.5 Flash, GPT-4.1, also have comparative performance. This is intuitive as these models have been trained on large-scale data and optimized for a wide range of tasks.
However, there are also some exceptions.
GPT-4o, despite being a large and powerful model, performs worse than InternVL3.5 38B and Qwen2.5-VL 72B.
Smaller models like InternVL3.5 4B also outperform larger ones like InternVL3.5 14B and Qwen2.5-VL 7B.
It is interesting that in the same model family, the performance does not always improve with scale.
This suggests that scaling up general VQA or CQA training does not necessarily improve the capability to extract chart data. We may not need very large models for chart data extraction.

\textbf{F3: Tailored data and training methods matter more.}
We found that general VQA benchmarks are not reliable indicators of performance on chart data extraction. The rankings of MLLMs on benchmarks such as MMBench~\cite{liu2024mmbench} do not align well with their performance on our task. This discrepancy underscores the need for specialized evaluation: chart data extraction requires precise geometric reasoning and fine-grained value estimation, which are capabilities that are not fully exercised in standard VQA or even typical CQA tasks. Combined with \textbf{F2}, this suggests that for chart data extraction, task-specific data and training strategies may offer a more cost-effective path to improving model performance.

\textbf{F4: Even smaller models can output correct table structure.} 7/11 models achieve over 97\% format success rate, indicating that they can understand and follow the output format requirements well. This aligns with our preliminary study and meets our expectations. By providing a CSV template in the prompt, we guide the model to generate the desired structure. This capability is crucial for practical deployment, where strict adherence to table structure is often necessary.
This also inspires us that, in future applications, users can provide a table to be filled with MLLMs for more controllable data extraction.

\textbf{F5: Significant imbalance exists in accuracy across chart types.}
As shown in~\autoref{fig:benchmark-type}, for most MLLMs, bar charts, line charts, and scatter plots, which use Cartesian coordinates and are commonly seen in training corpora, result in higher accuracy. In contrast, radar charts, which use polar coordinates and are relatively rare, remain the most difficult.
Interestingly, GPT-4.1 achieves the best score on pie and radar charts (4.59\% and 8.83\%, respectively) but performs worse on bar charts.
Such an imbalance may stem from the training data distribution, multi-task instruction tuning, or distillation processes.

\begin{figure}[t]
    \centering
    \includegraphics[width=\linewidth]{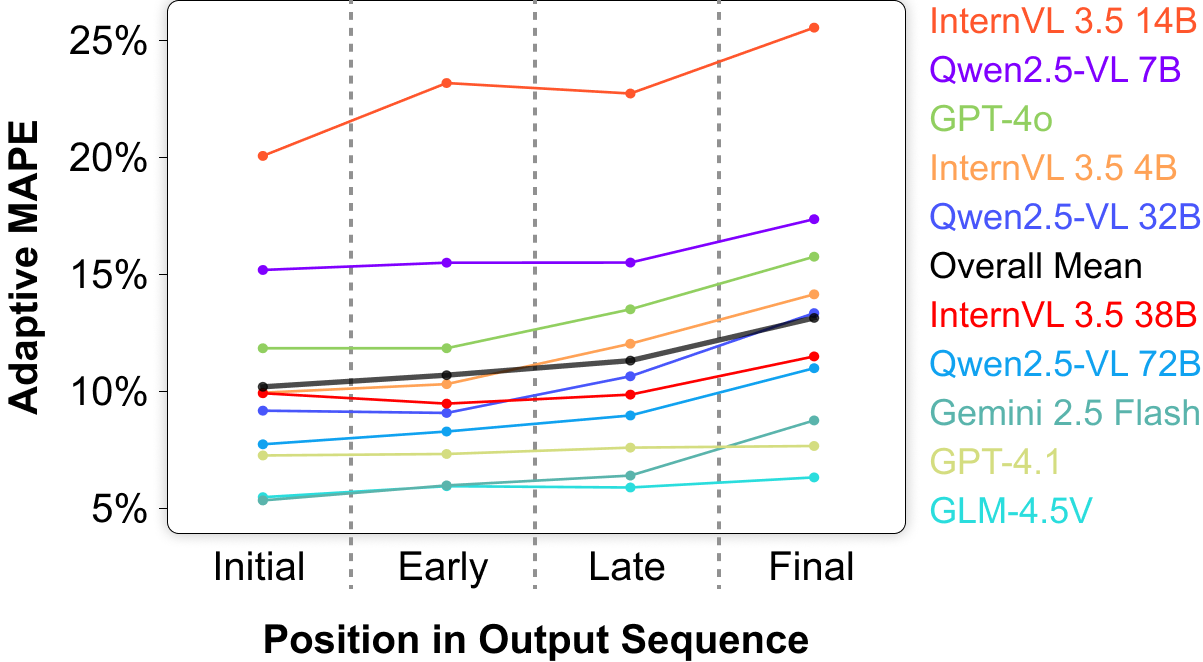}
    \caption{Trends in Adaptive MAPE relative to the position of values in the output sequence. ``Initial'' refers to the first 25\% of values in sequences, ``early'' to the next 25\%, ``late'' to the following 25\%, and ``final'' to the last 25\%.}
    \label{fig:trend}
    \Description{This figure shows the trends in Adaptive MAPE relative to the position of values in the output sequence. ``Initial'' refers to the first 25\% of values, ``early'' to the next 25\%, ``late'' to the following 25\%, and ``final'' to the last 25\%.}
\end{figure}

\textbf{F6: Table size has a limited impact on overall performance, but value accuracy degrades later in the output sequence.}
As shown in~\autoref{tab:benchmark-size}, format success rates are comparable in general across different table sizes, with minor variations between models. For Adaptive MAPE, some models perform slightly better on smaller tables, while others show improved accuracy on larger ones. Overall, smaller tables show a marginal advantage, but the effect is not substantial, suggesting that table size alone is not a strong determinant of performance.
However, when analyzing the Adaptive MAPE in relation to the position within the output sequence (\autoref{fig:trend}), a consistent trend emerges: values generated later in the sequence tend to have higher error. This indicates that while table size may not directly impact performance, the sequential nature of autoregressive text generation contributes to error accumulation, making later values less reliable.
Models may partially rely on earlier output values and the geometry relationships between corresponding visual marks to estimate subsequent values. Such dependency can be both beneficial and harmful.
Further exploration can be considered to mitigate this issue and improve robustness.

\begin{figure*}[t]
    \centering
    \includegraphics[width=\linewidth]{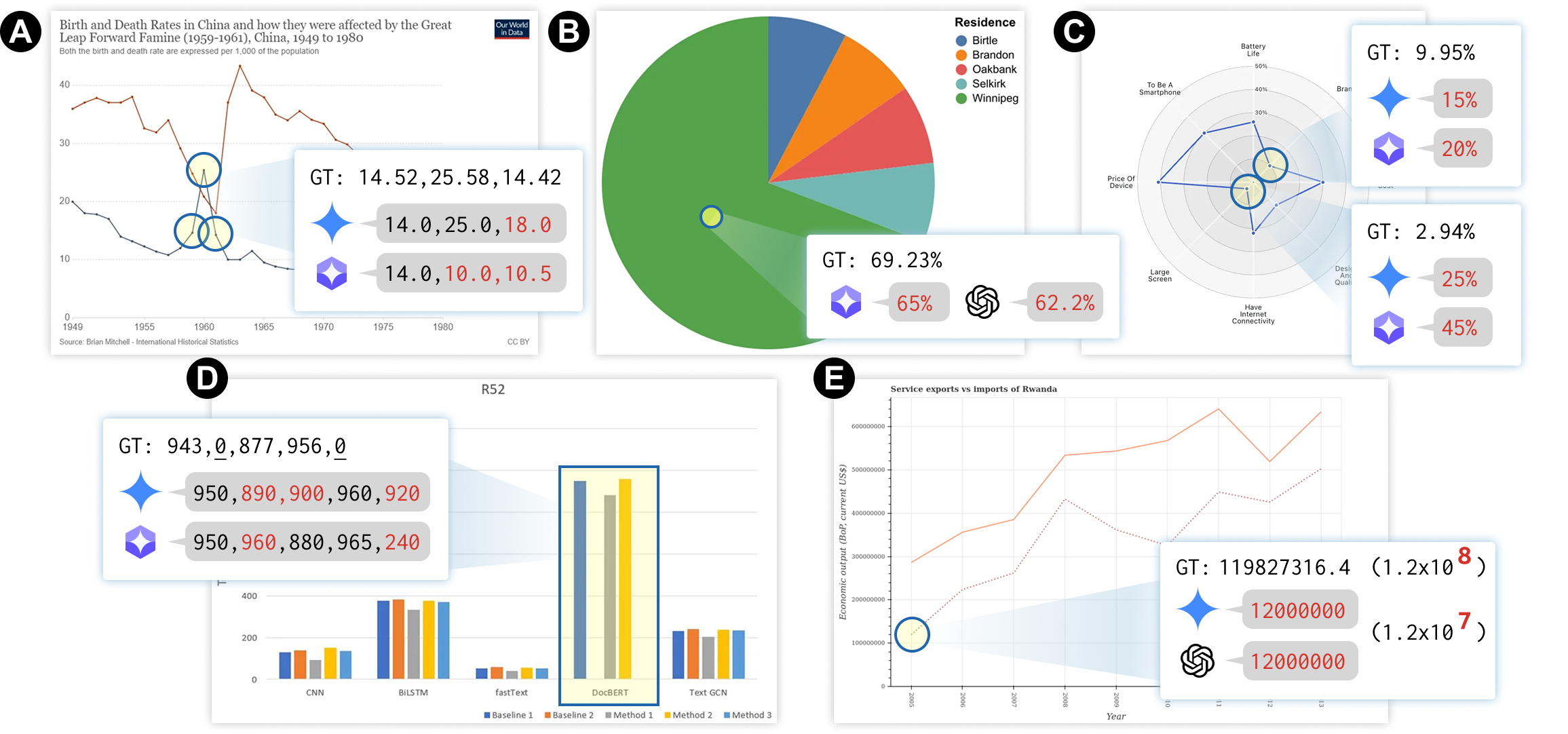}
    \caption{Common failure cases of MLLMs on our benchmark: (A) Sudden changes in value on line charts. (B) Pie slices exceeding 50\%. (C) Small values in radar charts. (D) Zero values in bar charts. (E) Large-magnitude axis ticks.}
    \label{fig:failure-case}
    \Description{This figure shows common failure cases of MLLMs on our benchmark: (A) Sudden changes in value on line charts. (B) Pie slices exceeding 50\%. (C) Small values in radar charts. (D) Zero values in bar charts. (E) Large-magnitude axis ticks.}
\end{figure*}

\begin{figure}[t]
    \centering
    \includegraphics[width=\linewidth]{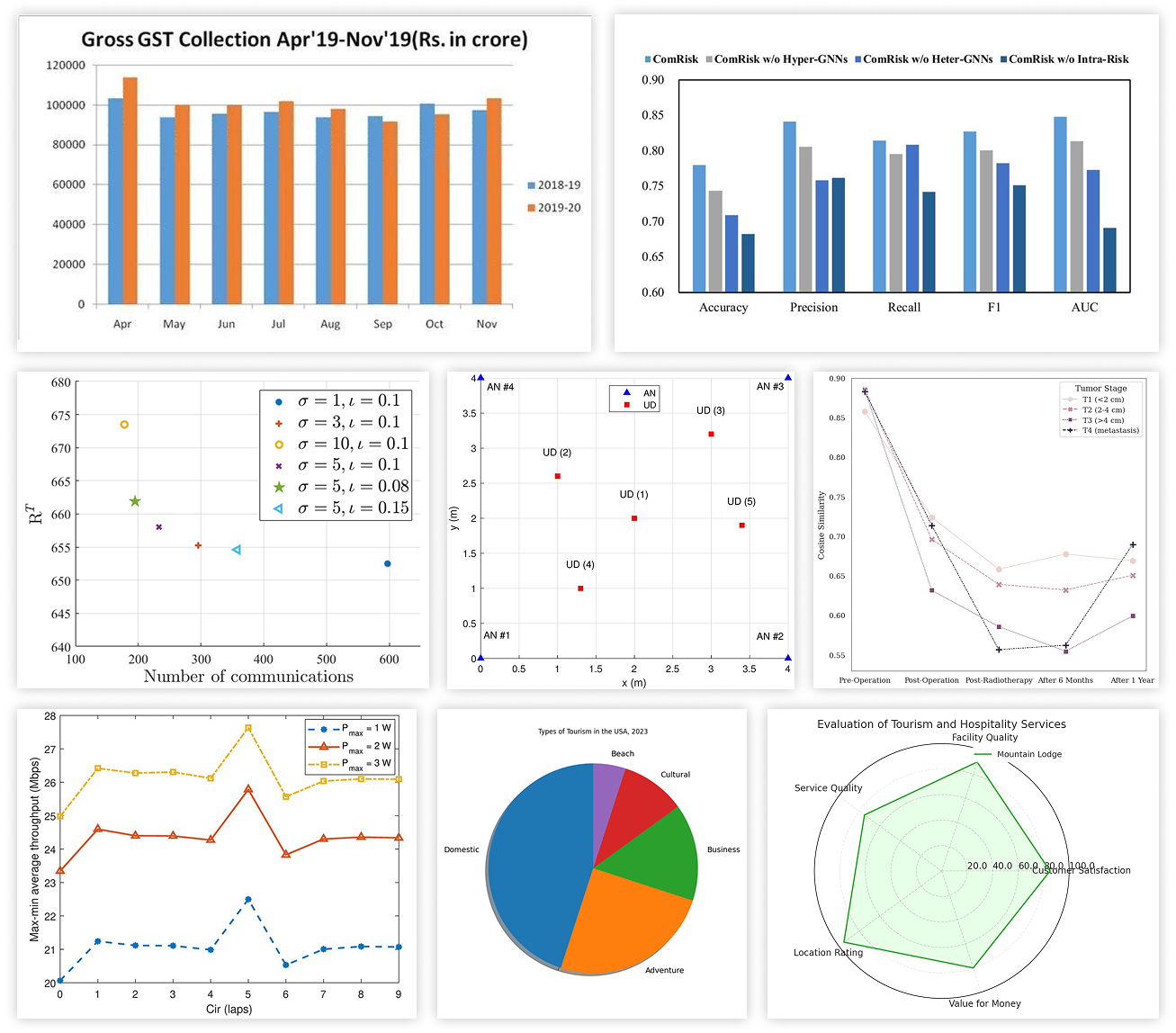}
    \caption{\revision{Success examples, where each chart achieves less than 1\% average Adaptive MAPE across Gemini 2.5 Flash, GLM-4.5V, and GPT-4.1.}}
    \label{fig:success-case}
    \Description{Success examples spanning diverse chart types (bar, scatter, line, pie, and radar). All charts achieve less than 1\% average Adaptive MAPE across Gemini 2.5 Flash, GLM-4.5V, and GPT-4.1.}
\end{figure}

\subsection{Analysis of Failure and Successful Cases} 

We analyzed the typical failure modes across models and identified several recurring patterns:

\textit{Sudden changes in value trends are often misinterpreted in line charts.} As shown in \autoref{fig:failure-case} (A), line charts with sharp spikes or drops frequently lead to inaccurate predictions. Models tend to repeat the same values or assume smoother trends, likely reflecting prior expectations learned during pretraining. This indicates a lack of robustness to abrupt visual transitions.

\textit{Models become less accurate when the slice in pie charts exceeds 50\%.} As shown in \autoref{fig:failure-case} (B), compared to a small slice, when a pie slice represents more than half of the chart, models are more likely to produce larger errors in estimating its value.
This may be due to a minority of large slices in training data, or the visual challenge of accurately judging angles and areas in such cases.

\textit{Small values near the origin are overestimated in radar charts.} As illustrated in \autoref{fig:failure-case} (C), values within the innermost grid rings, which represent the smallest magnitudes, are frequently inflated. This may stem from challenges in distinguishing fine-grained radial distances near the center, particularly in polar coordinate systems.

\textit{Zero-height bars often cause corruption in outputs.} As shown in \autoref{fig:failure-case} (D), when a bar chart includes bars with zero height, models frequently mess up the corresponding and following values.
This phenomenon is more obvious in stacked bar charts and grouped bar charts, as models tend to interpolate or replicate nearby values.
This suggests that models may struggle with discontinuities in visual encodings.
Training on more examples with zero values may help models better handle such cases.

\textit{Models are not good at counting the number of digits in large-magnitude values.} As shown in \autoref{fig:failure-case} (E), when axis ticks have large magnitudes (e.g., $>10^5$), models frequently miscount the number of digits, leading to errors in magnitude.
This is consistent with known weaknesses in LLMs regarding symbolic precision, similar to errors in character-counting tasks (e.g., ``how many r's in strawberry'').

\revision{Beyond the representative failure cases, we also present success examples in \autoref{fig:success-case} to illustrate the upper bound of current MLLM capabilities.
This contrast underscores the necessity of targeted training and interactive validation to address the long tail of challenging real-world charts.}
\begin{figure*}[h]
    \centering
    \includegraphics[width=0.95\textwidth]{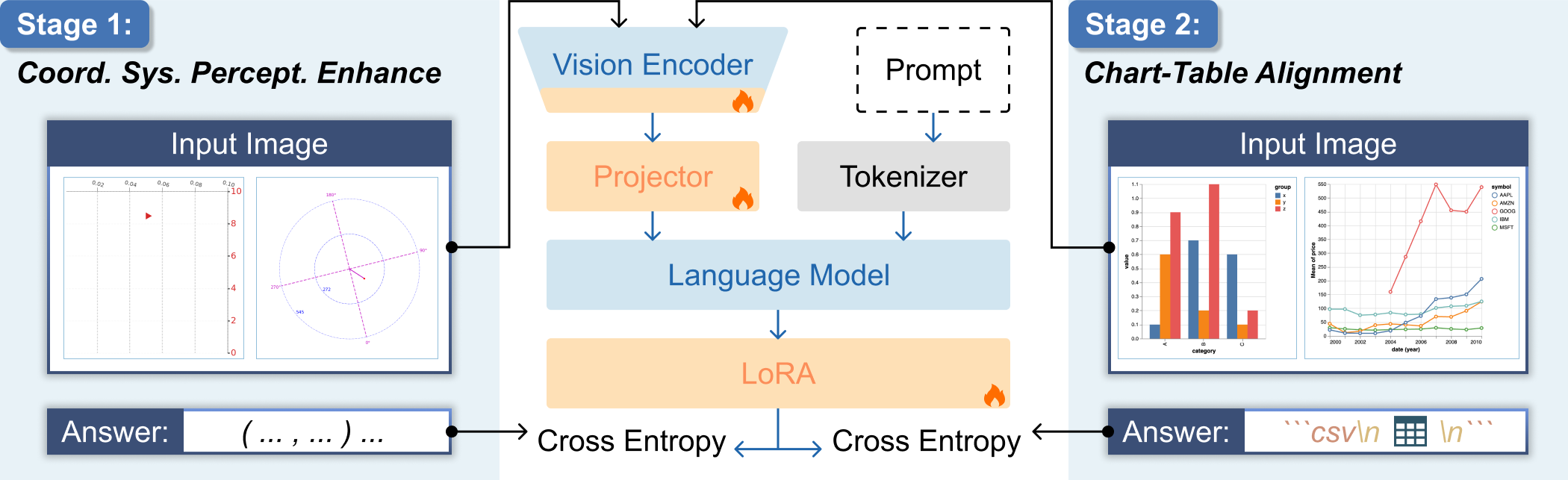}
    \caption{The proposed training framework includes a \textit{Coordinate System Perception Enhancement} stage and a \textit{Chart-Table Alignment} stage. During training, the final four layers of the vision encoder and the multimodal projector are fully fine-tuned, while the language model is adapted using LoRA.}
    \label{fig:training-framework}
    \Description{This figure shows the overall training framework, which includes a Coordinate System Perception Enhancement (CSPE) stage and a Chart-Table Alignment Enhancement (CTAE) stage.}
\end{figure*}

\section{\method{}}

Evaluation results on \benchmark{} indicate that existing MLLMs struggle with chart data extraction, particularly in achieving high numerical accuracy. While these models perform well on general vision-language tasks, they often lack the specialized capabilities needed to interpret chart-specific visual encodings.

Motivated by insights from visualization and HCI research on how people interpret or generate charts~\cite{shi2025chartist,viznet,10.1145/3290605.3300358}, we introduce \method{}, a two-stage training framework that mirrors the progressive learning process. Rather than directly aligning full charts with tables or code, we decompose the task into foundational perception and higher-level extraction.
The first stage, \textit{Coordinate System Perception Enhancement} (CSPE), is designed to improve the model's understanding of coordinate geometry and visual encodings.
The second stage, \textit{Chart-Table Alignment} (CTA), fine-tunes the model to generate structured data tables with high numerical fidelity.
An overview of the full framework is presented in~\autoref{fig:training-framework}.

\subsection{Coordinate System Perception Enhancement}

This stage aims to improve the model's understanding of coordinate geometry, enabling it to estimate the chart coordinates of individual visual marks. \textbf{F5} reveals an imbalanced performance across chart types: most MLLMs perform better on charts using Cartesian coordinates (e.g., bar and line charts) than on those using polar coordinates (e.g., pie and radar charts).

Intuitively, before a model can reliably recover data values from a chart, it must first grasp the coordinate system that governs the visual encodings. This reflects how humans learn to interpret charts, first understanding coordinate systems and then mapping visual cues to values. To facilitate this process, we design a pretraining task that explicitly targets coordinate system understanding.
Additionally, our analysis shows that models may rely on visual relationships between marks to infer values, leading to error accumulation in sequential outputs. By explicitly training models to reason about coordinate geometry, we aim to reduce this dependency and improve overall robustness.

In this task, for each type of coordinate system, the model is given images containing a single visual mark and must predict its coordinates. For Cartesian systems, the prediction takes the form of $(x, y)$; for polar systems, it is expressed as $(r, \theta)$. This task serves as an atomic subcomponent of chart data extraction, isolating the challenge of coordinate interpretation from the complexities of full-table generation.

\textbf{Training Objective.}
We formulate this task as a text generation problem and optimize it using cross-entropy loss. In this stage, the output could be just the coordinate pair, or it could also be output in CSV format. The example of Cartesian coordinates output in both formats is shown below:
\begin{center}
\begin{tabular}{|l|l|}
\ttfamily\textcolor{deepblue}{\textbf{Output Format 1:}} & \ttfamily\textcolor{deepblue}{\textbf{Output Format 2:}} \\
\ttfamily(23.5, 47.8) & \ttfamily\textasciigrave\textasciigrave\textasciigrave csv \\
& \ttfamily x,y\\
& \ttfamily 23.5,47.8\\
& \ttfamily\textasciigrave\textasciigrave\textasciigrave
\end{tabular}
\end{center}
To explore the impact of output format on this stage, we experiment with both formats in two-stage training and ablation studies and adopt the better one.

\textbf{Validation Metric.}
We evaluate performance using MAPE, treating each numeric value in the coordinate pair as an individual data point.

\begin{figure}[t]
    \centering
    \includegraphics[width=\linewidth]{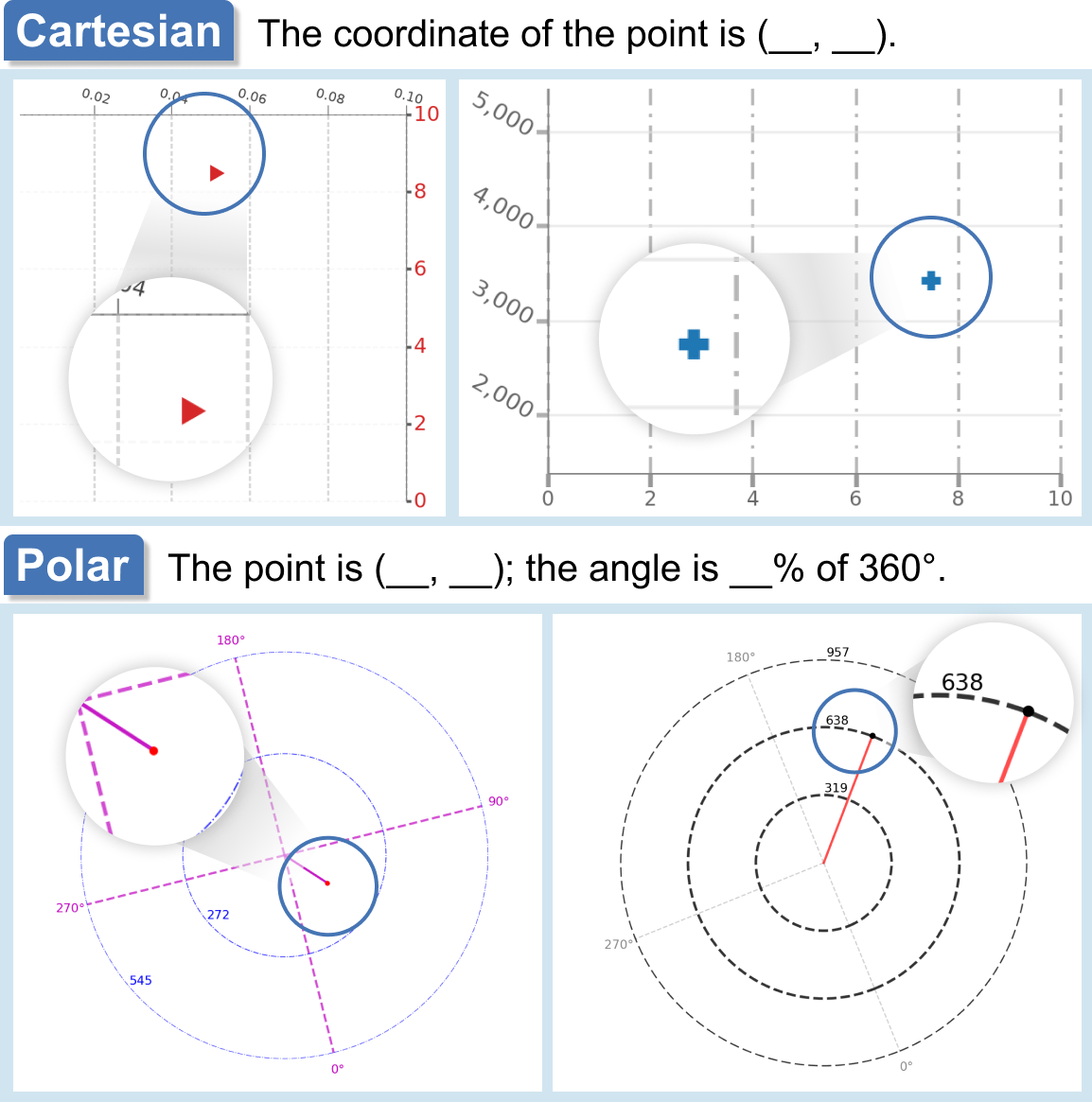}
    \caption{Examples of training samples for the Coordinate System Perception Enhancement stage.}
    \label{fig:stage1-example}
    \Description{This figure shows examples of training samples for the Coordinate System Perception Enhancement stage. The upper part illustrates examples of Cartesian coordinates, while the lower part presents examples of polar coordinates.}
\end{figure}

\textbf{Data Generation.}
We construct a large-scale synthetic dataset that includes both Cartesian and polar coordinate systems, consistent with the types of charts used in our benchmark.
For Cartesian coordinates, we use the Vega specifications to generate images, each containing a single point. The Axis ranges are randomly sampled to span a wide range of magnitudes, and point positions are randomly selected within these ranges. To promote style diversity, we vary Vega styling options such as axis placement, tick mark appearance, background color, point shape and size, and canvas resolution.

For polar coordinates, we use Matplotlib to generate single-point charts. We vary the reference direction (i.e., the $0^\circ$ axis) and the radius range. Points are placed at random angles between $0^\circ$ and $360^\circ$, with radii randomly sampled from the specified range. Chart appearance is further diversified using Matplotlib's style parameters. To simulate pie chart semantics, we additionally include the corresponding percentage (based on the angle) in the model output, ask the model to output $r$, $\theta$, and percentage.

This generation pipeline allows us to efficiently produce 40,000 samples for each type of coordinate system, and 1,000 samples for validation. For all samples, we also apply the same augmenting strategy as in the benchmark to simulate imperfection in real-world charts.
This pipeline can also be easily extended to include additional coordinate systems or visual styles as needed.
Examples of samples are illustrated in~\autoref{fig:stage1-example}.

\subsection{Chart-Table Alignment}
As with many vision-language tasks, fine-tuning on task-specific data is crucial for aligning the model with the intended objective. In our case, we fine-tune the model on chart-table pairs to improve its ability to extract structured data from visual representations. This stage builds upon the first: once the model has learned to interpret coordinate systems, it is further trained to recognize visual marks and translate them into accurate, structured table outputs.

\textbf{Training Objective.}
We optimize the model using cross-entropy loss as well. Given a chart and an instruction prompt, the model generates a predicted table in plain text. The training loss is computed between the predicted table and the ground truth table.

\textbf{Validation Metric.}
For validation, we use the same prompt template as defined in our benchmark, as shown in~\autoref{fig:benchmark-prompt}. After parsing the model's output into a structured table, we calculate the Adaptive MAPE between the predicted and ground-truth values. To ensure robust evaluation, individual errors are capped at 100\%, and any unparsable or missing values are also assigned a 100\% error.

\textbf{Data Generation.}
We employ the same generation pipeline as used in the synthetic portion of our benchmark to create chart-table pairs. Due to the limited availability of high-quality open-source chart datasets, we primarily rely on synthetic data generated using Vega and Matplotlib. We generate a total of 50,000 samples for training and 1000 samples for validation, covering a wide range of chart types, styles, and data distributions.

In this stage, the prompt used in training is slightly different from that used in validation. Instead of providing a partially completed CSV template to fill in, during training, the prompt instructs the model to generate the entire table from scratch within a code block.

\subsection{Model Configuration}

\textbf{Base Model.}
We use Qwen2.5-VL 7B as the base model for all experiments. This model is widely adopted for vision-language tasks and has demonstrated strong performance across a variety of VQA and CQA benchmarks.

\textbf{Training Setup.}
Our goal is to train the model to map visual elements (e.g., pixels) to numerical values based on the underlying chart coordinate system, capturing the correct slope, offset, and unit. This mapping is highly sensitive to visual features such as tick density, line width, anti-aliasing effects, font styles, dual-axis configurations, and polar layouts.
To enhance the model's ability to interpret such fine-grained visual signals, we fully fine-tune the last four layers of the vision encoder. These layers produce the visual features that are passed into the multimodal projector, and their norm scales and positional sensitivity determine how the language model interprets numerical structure.

The multimodal projector plays a critical role in aligning visual features with the language model's token space. Its output scale and variance directly impact numerical stability during generation. As we adapt both the tail of the vision encoder and the language model, the projector must be trained to maintain alignment in feature distributions, ensuring consistent unit interpretation across different font styles and axis ranges. Therefore, we also fully fine-tune the multimodal projector.

To ensure that the model outputs well-formatted tables, which preserve correct column order, row counts, grouping structure, and units, we adapt the language backbone using LoRA~\cite{hu2022lora}. Since this is primarily a formatting and control task, applying lightweight adaptation via LoRA is efficient and effective.

\subsection{Experiment Settings}

We conduct the experiments using the proposed two-stage training framework to evaluate its effectiveness in improving MLLM performance on chart data extraction tasks. The experimental design includes three configurations:
\begin{itemize}
\item \textbf{Full Framework.} Fine-tuning the base model with both the CSPE and CTA stages.
\item \textbf{Ablation 1.} Fine-tuning the base model with only the CTA stage (ablation of stage 1). This setting serves as a baseline to assess the contribution of the CSPE stage.
\item \textbf{Ablation 2.} Fine-tuning the base model with only the CSPE stage (ablation of stage 2). This setting isolates the effect of the CTA stage on overall performance.
\end{itemize}

\textbf{Parameter Settings.}
Experiments are conducted on a server equipped with 4 NVIDIA A100 80GB GPUs.
For the LoRA adapters, we use a rank of 8 and an alpha of 16 to adapt the language backbone.
Combined with the fully unfrozen vision layers and multimodal projector, this setup results in approximately 1.70\% of the total model parameters being trainable.
We use AdamW as the optimizer, with a weight decay of 0.1 and betas set to (0.9, 0.95) for both stages.
The learning rate is set to 1e-5 for the vision encoder, 5e-5 for the multimodal projector, and 1.5e-4 for the LoRA parameters.
We use a batch size of 4 with gradient accumulation steps set to 4. Each stage is trained for up to 10 epochs.

\textbf{Evaluation.}
We evaluate all trained models on the benchmark introduced earlier and report performance using Adaptive MAPE as the primary metric.

\subsection{Results}

The performance of different training configurations on our benchmark is summarized in~\autoref{tab:benchmark-type}.

\textbf{Compare to Existing MLLMs.}
After training with our full framework, the model achieves the lowest Adaptive MAPE of 4.87\%, significantly outperforming the base model Qwen2.5-VL 7B (16.01\%) as well as larger models such as GLM-4.5V (5.94\%) and Gemini 2.5 Flash (6.72\%).
Our model also shows consistent improvements across all chart types, with more balanced performance between Cartesian and polar coordinate systems. Notably, the Adaptive MAPE on radar charts improves substantially from 25.47\% to 3.21\%, indicating a stronger understanding of polar encodings after fine-tuning.
The Adaptive MAPE of scatter plots drops to below 1\% (0.85\%), which is expected since scatter plots are effectively the simplest form of Cartesian charts, and the CSPE stage includes scatter-like samples with a single point.
The format success rate also improves significantly from 89.58\% to 99.11\%, indicating better control over structured outputs. Considering the lightweight language backbone of Qwen2.5-VL 7B, this result is impressive.

\begin{table}[ht]
\caption{Comparison of our model, existing chart-specific models, and advanced MLLMs on RNSS and RMS.}
\label{tab:chart-specific}
\centering
\begin{tabular}{l|cc}
\toprule
\textbf{Model} & \textbf{RNSS} & \textbf{RMS} \\ \midrule
ChartInstruct & 7.69 & - \\
ChartAssistant & 19.59 & - \\
MatCha & 74.79 & 19.52 \\
TinyChart & 41.75 & 41.18\\
UniChart & 56.60 & 43.58 \\ \midrule
Gemini 2.5 Flash & 62.22 & 69.69 \\
GLM-4.5V & 71.45 & 73.47 \\ \midrule
Ours & \textbf{75.92} & \textbf{78.61} \\
\bottomrule
\end{tabular}%
\end{table}

We also compare our model with several chart-specific models, including ChartInstruct~\cite{masry2024chartinstruct}, ChartAssistant~\cite{meng2024chartassisstant}, UniChart~\cite{masry2023unichart}, TinyChart~\cite{zhang2024tinychart}, and MatCha~\cite{liu2023matcha}.
Because these models are not able to produce well-formed tables under our prompt design, we follow the prompting strategy used by Islam et al.~\cite{islam2024large} and evaluate their outputs using the RNSS~\cite{masry2022chartqa} and RMS~\cite{liu2023deplot} metrics.
The results are presented in~\autoref{tab:chart-specific}. For ChartInstruct and ChartAssistant, which fail to generate valid table structures but still output partial numeric values, we report only their RNSS scores.
Across both metrics, our model achieves the best performance among all chart-specific baselines.

\textbf{Ablation Study.}
Ablation results confirm that both stages of our framework contribute meaningfully to overall performance.
When the Coordinate System Perception Enhancement stage is removed, performance drops to 5.94\%, suggesting that fine-tuning solely on chart-table pairs is insufficient for learning coordinate-based reasoning. A foundational understanding of coordinate geometry is necessary for accurate value extraction.
Conversely, removing the Chart-Table Alignment stage leads to a performance of 11.17\%, indicating that learning to output structured tables is also critical.

These results validate the effectiveness of \method{} in enhancing the chart data extraction capabilities of MLLMs. By explicitly targeting coordinate perception and structured output generation, our approach enables models to achieve high numerical accuracy and robustness across diverse chart types. But it is also important to note that an Adaptive MAPE of 4.87\% does not imply that the model is already reliable enough to function as a fully automatic data extraction tool. Rather, this result shows that a lightweight 7B model, when trained with our progressive framework, can surpass large state-of-the-art MLLMs by a clear margin. This improvement demonstrates the potential of the approach: if substantial gains can be achieved at the 7B scale, further scaling in model capacity and training data may bring the error rate closer to the threshold required for standalone, reliability-critical applications.

At the current stage, the model alone cannot serve as a dependable extractor without human oversight. Following the design principles of mixed-initiative systems, a practical path toward reliability is to trade a small amount of efficiency for user verification and lightweight correction. In this paradigm, the model provides an initial extraction, and the human ensures correctness through targeted interaction. We then conduct a user study to assess whether our trained model can effectively support such a workflow.
\section{User Study}

In this study, we try to answer: \textit{Can MLLMs be integrated into an interactive workflow that enables reliable chart data extraction through user verification and correction?} Although our 7B MLLM outperforms top-tier commercial proprietary models, interactive support remains essential for reliable chart data extraction. Even the best-performing models still produce errors exceeding 4\%, which may not satisfy the accuracy requirements of certain analytical tasks. Moreover, due to the black-box nature of MLLMs, users cannot fully trust extracted values without verification.

Mixed-initiative workflows allow humans to validate and correct model outputs, avoiding the fragility of fully automatic pipelines. To evaluate whether an MLLM-assisted workflow is practically useful, we conducted a controlled user study\footnote{The study has been approved by State Key Lab of CAD\&CG, Zhejiang University.} focusing on both usability and efficiency. Our goal is to determine whether users, with adequate interface support, can efficiently validate model-extracted values at pixel precision and whether they perceive such a workflow as reliable and suitable for real-world use.

\begin{figure*}[ht]
    \centering
    \includegraphics[width=\textwidth]{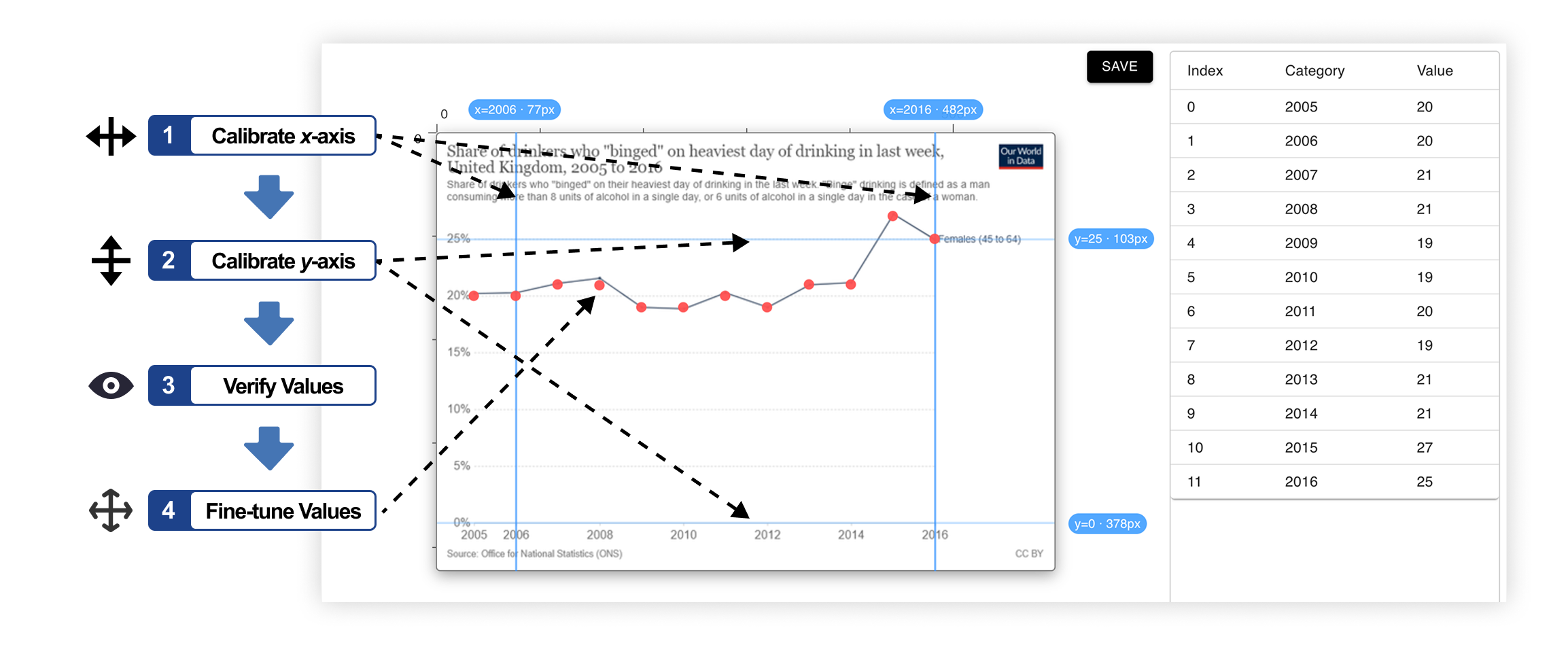}
    \caption{Using the prototype system to validate and correct extracted chart data for a basic line chart. Users first calibrate the chart; then all extracted data points are overlaid for verification. Users can drag any data point to correct its value if necessary.}
    \label{fig:interface}
    \Description{This figure shows the user interface of our prototype system for validating and correcting extracted chart data. It shows an example line chart with data points overlaid, and four calibration lines to calibrate the chart axes.}
\end{figure*}

\subsection{Prototype System}
To support verification and correction, the system must project model-extracted numerical values back onto the chart image. This requires establishing a mapping from chart coordinate space to image pixel space, enabling each value to be visualized at its corresponding position.
This is conceptually the inverse of what traditional extraction tools compute. Interactive tools such as WebPlotDigitizer first ask users to calibrate two points along the chart axes to establish the coordinate system. After calibration, users extract data by clicking on marks. We adopt the same calibration strategy.

Our prototype supports both Cartesian and polar coordinates. Users perform a brief calibration step, after which the system overlays model-extracted values onto the chart as movable markers. Users can click and drag these markers to correct errors, achieving pixel-level accuracy comparable to traditional tools. To simplify projection, we require the model to output tables in standardized formats for each chart type. \autoref{fig:interface} shows an example interface for basic line charts.

\subsection{Procedure}

We recruited 12 participants (A1---A12, male: 9, female: 3, age: 21---27) from a local university. All participants were familiar with common chart types such as bar, line, and pie charts. They were compensated according to our university's guidelines.
We sampled 24 real-world charts from our benchmark to better reflect practical scenarios. Twelve charts contained six data marks, and twelve contained fifteen. The set covered basic bar, stacked bar, grouped bar, basic line, grouped line, and pie charts.
Each session lasted approximately 45 minutes and included five phases:
\begin{itemize}
\item \textbf{Introduction (3 min)}. Participants were introduced to chart data extraction and the goal of obtaining accurate numerical values.
\item \textbf{System demonstration (8 min)}. We demonstrated calibration, verification, and correction using six example charts representing all chart types included in the task.
\item \textbf{Practice (5 min).} Participants familiarized themselves with the interface using sample charts.
\item \textbf{Extraction tasks ($\approx$20min)}. Each participant processed all 24 charts in randomized order. For each chart, participants validated every data point and corrected errors when necessary. They could only proceed after an experimenter confirmed pixel-level accuracy. We recorded the task completion time for each chart.
\item \textbf{Questionnaire and interview (8 min)}. Participants completed a SUS~\cite{brooke2013sus} questionnaire and three custom Likert-scale items, followed by a brief semi-structured interview.
\end{itemize}

\subsection{Measures and Analysis}

We employed a mixed set of quantitative and qualitative measures to evaluate how well the MLLM-based workflow supports reliable chart data extraction.

\textbf{Quantitative Measures.}
We collected task completion time for each chart and administered a post-study questionnaire consisting of the SUS and three five-point Likert items: \textbf{Q1}. I am satisfied with the extracted data after validation and correction. (This captures the perceived final quality of the workflow.) \textbf{Q2}. I perceive the model-extracted data as accurate. (This measures trust in raw model output before correction.) \textbf{Q3}. I would like to use this tool the next time I need to extract data from charts. (This evaluates acceptance and potential real-world adoption.)

\textbf{Qualitative Feedback.}
We also conducted brief semi-structured interviews focusing on (1) the ease of the calibration process, (2) the clarity of visual grounding during verification and correction.
For analysis, two researchers reviewed interview notes and grouped them into recurring themes through discussion.

\subsection{Results}

\textbf{Task Completion Time.}
On average, participants completed one chart with six data points in 31.35 seconds (SD = 8.39) and one chart with fifteen data points in 32.87 seconds (SD = 9.32).
To contextualize efficiency, we compared these results with the task times reported in Figure 7 of ChartSense. For bar charts, including grouped and stacked variants, our system enabled users to verify 15 data points in 31.73 seconds on average (SD=8.93), which is even faster than the approximately 34.5 seconds required for nine-point basic bar charts using ChartSense. Similarly, for basic line and grouped line, users completed fifteen-point verification in 36.31 seconds on average (SD=9.74), substantially faster than the approximately 57.5 seconds reported by ChartSense for only nine-point line charts.
These comparisons suggest that combining MLLM-assisted automation with interactive correction yields a verification efficiency that surpasses that of prior mixed-initiative approaches.

\textbf{System Usability.}
Participants rated the system highly on usability~\cite{bangor2008empirical}, with an average SUS score of 94.79 (SD=3.14).
The three Likert-scale questions yielded positive outcomes: \textit{Q1 (Final result satisfaction):} mean=4.92, SD=0.28; \textit{Q2 (Perceived model accuracy):} mean=3.67, SD=0.47; \textit{Q3 (Future use preference):} mean=5.00, SD=0.00.
The results show that participants were satisfied with the final corrected output (Q1), and all participants expressed willingness to use the system in the future (Q3). However, Q2 scores were noticeably lower, suggesting that although the workflow is effective, participants do not fully trust the model's raw outputs without verification. This confirms the need for interactive validation as part of model-assisted extraction workflows.

\textbf{Qualitative Themes.}
Analysis of the interview notes revealed two recurring themes related to the calibration, verification, and correction processes.

\textit{Calibration is easy to perform but remains the primary time cost.}
Most participants described the calibration step as intuitive and quick, yet still the dominant contributor to total task time. Several participants proposed improvements, such as enabling automatic detection of calibration points (A4) or allowing users to reuse calibration parameters for similar charts to avoid repeated setup (A8).

\textit{Visual grounding supports verification well, but can be refined.}
Participants generally found the overlay of extracted data points clear and helpful for checking accuracy. However, they also suggested enhancements. For instance, A2 proposed additional visual cues, such as small vertical lines indicating alignment with bar tops, to aid precision. A2 and A9 both recommended interaction techniques, such as snapping markers to the nearest visual boundary to ease fine adjustments.

\subsection{Summary}
The user study shows that an MLLM-assisted interactive workflow can effectively support reliable chart data extraction. Participants were able to verify and correct model-extracted values efficiently and rated the overall workflow highly on usability. At the same time, feedback indicates clear directions for interface refinement, particularly in improving calibration efficiency and enhancing visual grounding during verification.
Overall, these results demonstrate that MLLMs could also benefit from collaborating with humans to provide a promising approach to chart data extraction.
\section{Discussion}

In this section, we discuss the broader implications of our findings and potential extensions of our approach.

\subsection{Extending to Composite and Layered Charts}

While our current work focuses on charts with single mark types (e.g., bars, lines, or points), many real-world visualizations involve composite or layered encodings, such as bar charts with overlaid line charts, bubble charts that combine scatter and size encodings, or dual-axis plots. These charts introduce additional complexity in both coordinate system interpretation and mark-level disentanglement.
\method{} is naturally extensible to these more complex cases.
Given a composite chart type, we can identify its underlying coordinate system(s) and decompose the chart into atomic mark encodings. With this decomposition, we can generate corresponding training data for the CSPE stage. Then, CTA can be applied to train the model to extract structured data from the full chart, regardless of mark complexity.

\subsection{\revision{Implications for Human-AI Workflows}}

\revision{Our findings suggest integrating MLLMs into existing chart digitization tools, such as WebPlotDigitizer, to support a mixed-initiative workflow that combines automation with user verification. An MLLM can first generate an initial candidate data table directly from a chart image, which is then preloaded into the tool as editable values or plotted points. This removes the need for users to begin with an empty canvas and manually place control points.
In this setting, the interactive interface shifts from primary extraction to validation and correction. Tools like WebPlotDigitizer already support axis calibration and value adjustment, allowing users to identify and fix errors in ambiguous or unreliable regions. This creates a lightweight validation loop that fits naturally into existing practice: the model performs large-scale, repetitive extraction, while human effort is focused on judgment-intensive steps. The workflow reduces effort without sacrificing accuracy and keeps users in control of the final data. It positions MLLMs as complementary components that turn manual digitization tools into effective human-AI workflows.}

\subsection{Implications for General CQA}

Our findings also carry important implications for CQA. If a model cannot accurately extract underlying data values from a chart, its ability to answer quantitative questions may rely on superficial cues or spurious correlations rather than true chart understanding. This raises questions about the reliability of current CQA benchmarks, which often overlook data perception quality.
To address this, we introduce a label-free data extraction benchmark that offers a more targeted and interpretable evaluation of chart comprehension. By isolating the data perception component, our benchmark enables clearer diagnosis of whether errors stem from flawed perception or faulty reasoning.
Future CQA training and evaluation could benefit from decomposing the chart understanding ability into two aspects: (1) the perception ability responsible for interpreting the underlying data of the chart, and (2) the reasoning ability operating over known values. The former could be improved upon using the strategies proposed in this work, while the latter could leverage existing, high-performing reasoning LLMs.

\subsection{Future Work}

This study opens several promising directions for future research:

\revision{\textbf{Enhancing Chart Diversity.}
While the current benchmark already supports systematic comparison of chart data extraction performance across MLLMs and helps surface key failure modes, expanding it with more real-world charts and greater chart diversity would enable a more complete evaluation. For instance, it remains unclear how MLLMs behave on real-world charts affected by stronger distortions, such as low-quality screenshots or misaligned scans. Broader coverage of real-world charts would help expose weaknesses that are not well captured in partially synthetic settings. As future work, we plan to extend the benchmark with a wider range of real-world charts and apply additional augmentation techniques to simulate more severe distortions. In parallel, training models on real-world charts with more varied visual styles may improve robustness and help them handle unconventional or noisy design choices more reliably.}

\textbf{Exploring the Impact of Table Shape.}
We observed that value accuracy may correlate with the position of values in the output sequence. As a table can be transformed to multiple shapes, this motivates further investigation into how different table shapes, such as varying numbers of rows and columns or the presence of merged cells, affect model performance. Understanding these structural influences could inform the design of more effective table representations and extraction strategies.

\textbf{Incorporating Mechanistic Interpretability.}
As MLLM performance improves, interpretability becomes more important. Users will want to understand how reliable the values extracted by the model are.
Future work could involve integrating uncertainty estimation to provide confidence scores for individual predictions. This would help users assess the reliability of the extracted values. Additionally, exposing intermediate signals, such as attention distributions or predicted error likelihoods, could support automatic or interactive inference-time interventions~\cite{chen2025spatial,liu2025reducing}, enabling users to adjust model outputs when uncertainty is high.
\section{Conclusion}

In this work, we investigate whether MLLMs can serve as reliable chart data extractors.
We introduce a comprehensive benchmark that evaluates chart data extraction across diverse chart types and visual styles. Using this benchmark, we systematically assess advanced MLLMs and find that, although they can generate well-structured tables, they still struggle with numerical accuracy.
To address this challenge, we propose a two-stage training framework that mirrors the human chart-reading process, enabling models to learn chart data extraction progressively. Our approach leads to substantial improvements, achieving state-of-the-art performance on our benchmark.
A user study further validates the effectiveness of our method, demonstrating that, when paired with human verification, our model supports accurate and efficient chart data extraction in practical scenarios.
Overall, this work provides a detailed examination of current MLLM limitations and presents an effective framework for improvement, laying the groundwork for future research on reliable chart data extraction.

\begin{acks}
The work was supported by Zhejiang Provincial Natural Science Foundation of China under Grant No. LD25F020003, NSFC (62421003, 62402428), and Ningbo Yongjiang Talent Programme (2023A-396-G). The authors gratefully acknowledge the support of Zhejiang University Education Foundation Qizhen Scholar Foundation.
\end{acks}

\bibliographystyle{ACM-Reference-Format}
\bibliography{main}

\appendix

\end{document}